\documentclass[amstex,12pt,thmsa,sw20bams]{article}
\usepackage{amsfonts}
\usepackage[utf8]{inputenc}
\usepackage{graphicx}
\usepackage[francais]{babel}
\usepackage{amssymb}
\usepackage{amsfonts}
\usepackage{amsmath}
\usepackage{caption}
\usepackage{subcaption}

\input tcilatex
\begin{document}

\author{Vartan Choulakian and Jacques Allard \and Universit\'{e} de Moncton,
Canada \and email: vartan.choulakian@umoncton.ca, jacques.allard@umoncton.ca}
\title{Uncovering and displaying the coherent groups of rank data by
exploratory riffle shuffling }
\date{November 2020}
\maketitle

\begin{abstract}
Let $n$ respondents rank order $d$ items, and suppose that $d<<n$. Our main
task is to uncover and display the structure of the observed rank data by an
exploratory riffle shuffling procedure which sequentially decomposes the n
voters into a finite number of coherent groups plus a noisy group: where the
noisy group represents the outlier voters and each coherent group is
composed of a finite number of coherent clusters. We consider exploratory
riffle shuffling of a set of items to be equivalent to optimal two blocks
seriation of the items with crossing of some scores between the two blocks.
A riffle shuffled coherent cluster of voters within its coherent group is
essentially characterized by the following facts: a) Voters have identical
first TCA factor score, where TCA designates taxicab correspondence
analysis, an L$_{1}$ variant of correspondence analysis; b) Any preference
is easily interpreted as riffle shuffling of its items; c) The nature of
different riffle shuffling of items can be seen in the structure of the
contingency table of the first-order marginals constructed from the Borda
scorings of the voters; d) The first TCA factor scores of the items of a
coherent cluster are interpreted as Borda scale of the items. We also
introduce a crossing index, which measures the extent of crossing of scores
of voters between the two blocks seriation of the items. The novel approach
is explained on the benchmarking SUSHI data set, where we show that this
data set has a very simple structure, which can also be communicated in a
tabular form.


\end{abstract}

\section{\textbf{Introduction}}

Ordering the elements of a set is a common decision making activity, such
as, voting for a political candidate, choosing a consumer product, etc. So
there is a huge literature concerning the analysis and interpretation of
preference data scattered in different disciplines. Often rank data is
heterogenous: it is composed of a finite mixture of components. The
traditional methods of finding mixture components of rank data are mostly
based on parametric probability models, distance or latent class models, and
are useful for \textit{sparse} data and not for \textit{diffuse} data.

Rank data are sparse if there are at most a small finite number of
permutations that capture the majority of the preferences; otherwise they
are diffuse. As a running example in this paper, we will consider the famous
benchmarking SUSHI data set enumerating $n=5000$ preferences of $d=10$
sushis, see $\left[ 1\right] $. The SUSHI data set is diffuse, because there
are at most three counts for one observed permutation. It has been analyzed,
among others by $\left[ 2,3,4\right] $.

A second data set that we shall also analyze is the APA dataset of size $%
n=5738$ by $d=5$, see $\left[ 5\right] $. APA data set is also considered as
non sparse by $\left[ 2\right] $, because all the 120 permutations exist
with positive probability.

For a general background on statistical methods for rank data, see the
excellent monograph by $\left[ 6\right] $ and the book $\left[ 7\right] $.

\subsection{Riffle shuffle}

The riffle shuffle, see $\left[ 8\right] $, is considered the most popular
method of card shuffling, in which one cuts a deck of $d$ cards (aka items)
into two piles of sizes $d_{1}$ and $d_{2}$, respectively, and successively
drops the cards, one by one, so that the piles are interleaved into one deck
again.

Let $V,$ named a voting profile, be a set of $n$ preferences on $d$ items.
Based on riffle shuffling ideas, $\left[ 2\right] $ proposed the notion of
riffled independence to model the joint probability distribution of observed
preferences of $V$. Independently, $\left[ 9\right] $ used it for visual
exploration of $V$, naming it two blocks partition of the Borda scored items
with crossing of some scores; this will be further developed here under the
following important\bigskip

\textbf{Assumption: }$d<<n.$ This means that the sample size $n$ is quite
large compared to the number of items $d$. $\bigskip $

SUSHI and APA data sets satisfy this assumption.

The most important first step in the application of a riffle shuffling
procedure is how to partition the $d$ items into two disjoint subsets. In
the probabilistic riffle shuffling approach of $\left[ 2\right] $, the
partitioning step is essentially done using some adhoc approach in the case
of the SUSHI\ data set or based on background second order information of
the items in the case of the APA data set. While in the exploratory riffle
shuffling approach of this paper an optimal partition is obtained by
maximizing the cut norm of row centered data, or equivalently by taxicab
correspondence analysis of nega coded data.

We compare the two formulations of riffle shuffle, probabilistic and
exploratory, in section 10.

\subsection{Aim}

Our aim is to explore and display a given voting profile $V$ by sequentially
partitioning it into $G$ coherent groups plus a noisy group; that is,
\begin{equation}
V=\cup _{g=1}^{G}cohG(g)\cup noisyG,  \tag{1}
\end{equation}%
where $G$ represents the number of coherent groups and $cohG(g)$ is the $g$%
th coherent group. Furthermore, each coherent group is partitioned into a
finite number of coherent clusters; that is,
\begin{equation}
cohG(g)=\cup _{\alpha =1}^{c_{g}}cohC_{g}(\alpha )\text{ \ for }g=1,...,G,
\tag{2}
\end{equation}%
where $c_{g}$ represents the number of coherent clusters in the $g$th
coherent group. So the coherent clusters are the building blocks for the
coherent groups. We note the following facts:

Fact 1: The assumption $d<<n$ induces the new notion of coherency for the
clusters and consequently for the groups; it is a stronger characterization
than the notion of interpretability for groups as discussed in $\left[ 9%
\right] $.

Fact 2: Each coherent group and its clusters have the same latent variable
summarized by the Borda scale.

Fact 3: Given that the proposed method sequentially peels the data like
Occam's razor, the number of groups $G$ is calculated automatically.
Furthermore, outliers or uninformative voters belonging to the $noisyG$ are
easily tagged.

Fact 4: The approach is exploratory, visual, data analytic and is developed
within the framework of taxicab correspondence analysis (TCA). TCA is an L$%
_{1}$ variant of correspondence analysis developed by $\left[ 10\right] $.
TCA is a dimension reduction technique similar to principal component
analysis. In this paper, we shall use \textit{only} the first TCA factor
scores of the items and of the voters.

Two major advantages of our method are: First, we can easily identify
outliers. For the SUSHI\ data, our method tags 12.36\% of the voters as
outliers, which form the noisy group. While no outliers in the SUSHI\ data
have been identified in $\left[ 3,\ 4\right] $. Second, it provides a
tabular summary of the preferences which compose a coherent group. For
instance, consider the first mixture component of the SUSHI data given in $%
\left[ 4\right] $, where the modal ordering is almost the same as the Borda
scale ordering of the ten sushis in cohG(1) obtained by our method, see
Table 14 in this paper. The sample size of their first mixture component is
27.56 \%, which is much smaller than 48.36\%, the sample size of our
cohG(1), see Table 14. However, Table 13 of this paper provides a
tabular-visual summary of the 2418 preferences which form cohG(1). The
visual summary describes different kinds of equivalent similar riffle
shufflings of the 2418 preferences, and it provides further insight into the
structure of the data. Such interesting visual summaries are missing in $%
\left[ 3,\ 4\right] $.

\subsection{Highlights of a coherent cluster}

A \textit{coherent cluster} of voters has interesting mathematical
properties and is essentially characterized by the following facts:

a) Voters have identical unique first TCA factor score.

b) Any voter preference is easily interpreted as a particular riffle
shuffling of its items.

c) The nature of riffle shuffling of the items can be observed in the
structure of the contingency table of the first-order marginals constructed
from the Borda scorings of the voters belonging to the coherent cluster.

d) The first TCA factor scores of the items of a coherent cluster are
interpreted as Borda scale of the items.

e) We also introduce the crossing index, which measures the extent of
interleaving or the crossing of scores of voters between two blocks
seriation of the items in a coherent cluster.

\subsection{Organization}

This paper has eleven sections and its contents are organized as follows:
Section 2 presents an overview of TCA; section 3 presents some preliminaries
on the Borda coding of the data and related tables and concepts; section 4
presents Theorem 1, which shows that the first principal dimension of TCA
clusters the voters into a finite number of clusters; section 5 discusses
coherent clusters and their mathematical properties; section 6 discusses
riffle shuffling in a coherent cluster; section 7 introduces the crossing
index; section 8 introduces the coherent groups; section 9 presents the
analysis of APA data set; section 10 presents a comparison of the two
formulations of riffle shuffle probabilistic and exploratory; and finally we
conclude in section 11.

All mathematical proofs are relegated to the appendix. Details of the
computation are shown only for the first coherent group of SUSHI data set.

\section{An\ overview\ of\ taxicab\ correspondence\ analysis}

Consider a $n\times p$ matrix $\mathbf{X}$\textbf{\ }where $X_{ij}\geq 0.$
We have $\sum_{j=1}^{p}\sum_{i=1}^{n}\mathbf{X}_{ij}=X_{\ast \ast }.$ Let $%
\mathbf{P=X/}X_{\ast \ast }$ be the correspondence matrix associated to
\textbf{X}; and as usual, we define $p_{i\ast }=\sum_{j=1}^{p}p_{ij}$, $%
p_{\ast j}=\sum_{i=1}^{n}p_{ij}$. Let $\mathbf{D}_{n}=Diag(p_{i\ast })$ a
diagonal matrix with diagonal elements $p_{i\ast }$. Similarly $\mathbf{D}%
_{p}=Diag(p_{\ast j})$. Let $k=rank(\mathbf{P)}-1$\textbf{.}

In TCA the calculation of the dispersion measures $(\delta _{\alpha })$,
principal axes ($\mathbf{u}_{\alpha },\mathbf{v}_{\alpha }),$ principal
basic vectors $(\mathbf{a}_{\alpha },\mathbf{b}_{\alpha }),$ and principal
factor scores $(\mathbf{f}_{\alpha },\mathbf{g}_{\alpha })$ for $\alpha
=1,...,k$ is done in a stepwise manner. We put $\mathbf{P}%
_{1}=(p_{ij}^{(1)}=p_{ij}-p_{i\ast }\ p_{\ast j})$. Let $\mathbf{P_{\alpha }}
$ be the residual correspondence matrix at the $\alpha $-th iteration.

The variational definitions of the TCA at the $\alpha $-th iteration are

\begin{eqnarray*}
\delta _{\alpha } &=&\max_{\mathbf{u\in
\mathbb{R}
}^{p}}\frac{\left\vert \left\vert \mathbf{P_{\alpha }u}\right\vert
\right\vert _{1}}{\left\vert \left\vert \mathbf{u}\right\vert \right\vert
_{\infty }}=\max_{\mathbf{v\in
\mathbb{R}
}^{n}}\ \frac{\left\vert \left\vert \mathbf{P_{\alpha }^{\prime }v}%
\right\vert \right\vert _{1}}{\left\vert \left\vert \mathbf{v}\right\vert
\right\vert _{\infty }}=\max_{\mathbf{u\in
\mathbb{R}
}^{p},\mathbf{v\in
\mathbb{R}
}^{n}}\frac{\mathbf{v}^{\prime }\mathbf{P_{\alpha }u}}{\left\vert \left\vert
\mathbf{u}\right\vert \right\vert _{\infty }\left\vert \left\vert \mathbf{v}%
\right\vert \right\vert _{\infty }}, \\
&=&\max ||\mathbf{P_{\alpha }u||}_{1}\ \ \text{subject to }\mathbf{u}\in
\left\{ -1,+1\right\} ^{p}, \\
&=&\max ||\mathbf{P_{\alpha }^{\prime }v||}_{1}\ \ \text{subject to }\mathbf{%
v}\in \left\{ -1,+1\right\} ^{n}, \\
&=&\max \mathbf{v}^{\prime }\mathbf{P_{\alpha }u}\text{ \ subject to \ }%
\mathbf{u}\in \left\{ -1,+1\right\} ^{p},\mathbf{v}\in \left\{ -1,+1\right\}
^{n}.
\end{eqnarray*}%
The $\alpha $-th principal axes are%
\begin{equation}
\mathbf{u}_{\alpha }\ =\arg \max_{\mathbf{u}\in \left\{ -1,+1\right\}
^{p}}\left\vert \left\vert \mathbf{P_{\alpha }u}\right\vert \right\vert _{1}%
\text{ \ \ and \ \ }\mathbf{v}_{\alpha }\ =\arg \max_{\mathbf{v}\in \left\{
-1,+1\right\} ^{n}}\left\vert \left\vert \mathbf{P_{\alpha }^{\prime }v}%
\right\vert \right\vert _{1}\text{,}  \tag{3}
\end{equation}%
and the $\alpha $-th basic principal vectors are
\begin{equation}
\mathbf{a}_{\alpha }=\mathbf{P_{\alpha }u}_{\alpha }\text{ \ and \ }\mathbf{b%
}_{\alpha }=\mathbf{P_{\alpha }^{\prime }v}_{\alpha },  \tag{4}
\end{equation}%
and the $\alpha $-th principal factor scores are
\begin{equation}
\mathbf{f}_{\alpha }=\mathbf{D}_{n}^{-1}\mathbf{a}_{\alpha }\text{ \ and \ }%
\mathbf{g}_{\alpha }=\mathbf{D}_{p}^{-1}\mathbf{b}_{\alpha };  \tag{5}
\end{equation}%
furthermore the following relations are also useful%
\begin{equation}
\mathbf{u}_{\alpha }=sgn(\mathbf{b}_{\alpha })=sgn(\mathbf{g}_{\alpha })%
\text{ \ and \ }\mathbf{v}_{\alpha }=sgn(\mathbf{a}_{\alpha })=sgn(\mathbf{f}%
_{\alpha }),  \tag{6}
\end{equation}%
where $sgn(.)$ is the coordinatewise sign function, $sgn(x)=1$ \ if \ $x>0,$
\ and \ $sgn(x)=-1$ \ if \ $x\leq 0.$
The $\alpha $-th taxicab dispersion measure $\delta _{\alpha }$ can be
represented in many different ways%

\begin{equation}
\begin{array}{cccc}
\delta _{\alpha }\ &=&\left\vert \left\vert \mathbf{P_{\alpha }u}_{\alpha
}\right\vert \right\vert _{1}=\left\vert \left\vert \mathbf{a}_{\alpha
}\right\vert \right\vert _{1}=\mathbf{a}_{\alpha }^{\prime }\mathbf{v}%
_{\alpha }=\left\vert \left\vert \mathbf{D}_{n}\mathbf{f}_{\alpha
}\right\vert \right\vert _{1}=\mathbf{u}_{\alpha }^{\prime }\mathbf{D}_{n}%
\mathbf{f}_{\alpha }, & \tag{7}  \\
  &=&\left\vert \left\vert \mathbf{P_{\alpha }^{\prime }v}_{\alpha
}\right\vert \right\vert _{1}=\left\vert \left\vert \mathbf{b}_{\alpha
}\right\vert \right\vert _{1}=\mathbf{b}_{\alpha }^{\prime }\mathbf{u}%
_{\alpha }=\left\vert \left\vert \mathbf{D}_{p}\mathbf{g}_{\alpha
}\right\vert \right\vert _{1}=\mathbf{v}_{\alpha }^{\prime }\mathbf{D}_{p}%
\mathbf{g}_{\alpha }.&  
\end{array}
\end{equation}%
The $(\alpha +1)$-th residual correspondence matrix is
\begin{equation}
\mathbf{P_{\alpha +1}}=\mathbf{P_{\alpha }-D}_{n}\mathbf{f}_{\alpha }\mathbf{%
g}_{\alpha }^{^{\prime }}\mathbf{D}_{p}/\delta _{\alpha }. \tag{8}
\end{equation}%
An interpretation of the term $\mathbf{D}_{n}\mathbf{g}_{\alpha }\mathbf{f}%
_{\alpha }^{^{\prime }}\mathbf{D}_{p}/\delta _{\alpha }$ in (8) is that, it
represents the best rank-1 approximation of the residual correspondence
matrix $\mathbf{P_{\alpha }}$, in the sense of taxicab norm.

In CA and TCA, the principal factor scores are centered; that is,%
\begin{equation}
\sum_{i=1}^{n}f_{\alpha }(i)p_{i\ast }=0=\sum_{j=1}^{p}g_{\alpha }(j)p_{\ast
j}\text{ \ \ \ for \ \ }\alpha =1,...,k.  \tag{9}
\end{equation}

The reconstitution formula in TCA and CA is

\begin{equation}
p_{ij}=p_{i.}p_{.j}\left[ 1+\sum_{\alpha =1}^{k}f_{\alpha }(i)g_{\alpha
}(j)/\delta _{\alpha }\right] .  \tag{10}
\end{equation}

In TCA, the calculation of the principal component weights, $\mathbf{u}%
_{\alpha }$ and $\mathbf{v}_{\alpha },$ and the principal factor scores, $%
\mathbf{g}_{\alpha }$\ and \ $\mathbf{f}_{\alpha },$ can be accomplished by
two algorithms. The first one is based on complete enumeration based on
equation (3). The second one is based on iterating the transition formulae
(4,5,6). This is an ascent algorithm; that is, it increases the value of the
objective function at each iteration, see $\left[ 11\right] $. The iterative
algorithm could converge to a local maximum; so it should be restarted from
several initial configurations. The rows or the columns of the data can be
used as starting values.

The TCA map is obtained by plotting $(\mathbf{g}_{1},\mathbf{g}_{2})$ or $(%
\mathbf{f}_{1},\mathbf{f}_{2}).$

\section{Preliminaries}

In this section we review a) The Borda scoring of a voting profile \textbf{V}
into \textbf{R} and the Borda scale; b) Contingency table of the first order
marginals of \textbf{R}; c) The coded tables \textbf{R}$_{double}$ and
\textbf{R}$_{nega}.$

\subsection{Borda scorings and Borda scale}

Let $A=\{a_{1},a_{2},\ldots ,a_{d}\}$ denote a set of $d$
alternatives/candidates/items, and $V$ a set of $n$
voters/individuals/judges. In this paper we consider the linear
orderings/rankings/preferences, in which all $d$ objects are rank-ordered
according to their levels of desirability by the $n$ voters. We denote a
linear order by a sequence $\mathbf{s}=(a_{k_{1}}\succ a_{k_{2}}\succ \ldots
\succ a_{k_{d}})$, where $a_{k_{1}}\succ a_{k_{2}}$ means that the
alternative $a_{k_{1}}$ is preferred to the alternative $a_{k_{2}}.$ The
Borda scoring of $\mathbf{s}$, see $\left[ 12\right] ,$ is the vector $b(%
\mathbf{s)}$ where to the element $a_{k_{j}}$the score of $(d-j)$ is
assigned, because $a_{k_{j}}$ is preferred to $(d-j)$ other alternatives; or
equivalently it is the $j$th most preferred alternative. Let $\mathbf{R=(}%
r_{ij})$ be the matrix having $n$ rows and $d$ columns, where $r_{ij}$
designates the Borda score of the $i$th voter's preference of the $j$th
alternative. We note that the $i$th row of $\mathbf{R}$ will be an element
of $S_{d}$ the set of permutations of the elements of the set $\left\{
0,1,2,...,d-1\right\} .$\ A toy example of $\mathbf{R}$ is presented in
Table 1 for $n=4$ and $d=3$.

The Borda scale of the elements of $A$ is $\mathbf{\beta }=\mathbf{1}%
_{n}^{\prime }\mathbf{R}/n,$ where $\mathbf{1}_{n}$ is a column vector of $1$%
's having $n$ coordinates. The Borda scale seriates/orders the $d$ items of
the set $A$ according to their average scores: $\mathbf{\beta }(j)>\mathbf{%
\beta }(i)$ means item $j$ is preferred to item $i$, and $\mathbf{\beta }(j)=%
\mathbf{\beta }(i)$ means both items $(a_{i},a_{j})$ are equally preferred.
In the toy example of Table 1, the Borda scale seriates $\{A,B\}\succ C$.

Similarly, we define the reverse Borda score of $\mathbf{s}$ to be the
vector $\overline{b}$($\mathbf{s)}$, which assigns to the element $a_{k_{j}}$%
the score of $(j-1).$ We denote $\overline{\mathbf{R}}\mathbf{=(}\overline{r}%
_{ij})$ to be the matrix having $n$ rows and $d$ columns, where $\overline{r}%
_{ij}$ designates the reverse Borda score of the $i$th judge's nonpreference
of the $j$th alternative. The reverse Borda scale of the $d$ items is $%
\overline{\mathbf{\beta }}=\mathbf{1}_{n}^{\prime }\overline{\mathbf{R}}/n.$

We note that
\begin{equation*}
\mathbf{R+}\overline{\mathbf{R}}=(d-1)\mathbf{1}_{n}\mathbf{1}_{d}^{\prime }
\end{equation*}%
and
\begin{equation*}
\mathbf{\beta +}\overline{\mathbf{\beta }}=(d-1)\mathbf{1}_{d}^{\prime }.
\end{equation*}

\begin{tabular}{|l|crr|rrr|}
\multicolumn{7}{l}{\textbf{Table 1: Toy example with }$n=4$\textbf{\
preferences of }$d=3$ items\textbf{.}} \\
\hline
\multicolumn{1}{|r|}{} & $\mathbf{R}$&\multicolumn{1}{r}{}   & \multicolumn{1}{r|}{}  &\multicolumn{1}{r}{}&{$\overline{\mathbf{R}}$}&\multicolumn{1}{r|}{}\\
\hline
$A\succ B\succ C$ & \multicolumn{1}{|r}{0} & 1 & 2 & 2 & 1 & 0 \\
$A\succ C\succ B$ & \multicolumn{1}{|r}{1} & 0 & 2 & 1 & 2 & 0 \\
$B\succ A\succ C$ & \multicolumn{1}{|r}{0} & 2 & 1 & 2 & 0 & 1 \\
$B\succ C\succ A$ & \multicolumn{1}{|r}{1} & 2 & 0 & 1 & 0 & 2 \\
\hline\hline
Borda scale\ $\mathbf{\beta }$ & \multicolumn{1}{|r}{0.5} & 1.25 & 1.25 &  &
&  \\ \hline
$\text{reverse Borda scale}\overline{\text{ }\mathbf{\beta }}$ &
\multicolumn{1}{|l}{} &  &  & 1.5 & 0.75 & 0.75 \\ \hline
\textbf{nega\ }$n\overline{\mathbf{\beta }}$ & \multicolumn{1}{|l}{} &  &  &
6 & 3 & 3 \\ \hline
\end{tabular}

\subsection{Contingency table of first-order marginals}

The contingency table of first order marginals of an observed voting profile
$V$ on $d$ items is a square $d\times d$ matrix \textbf{M}, where $\mathbf{M(%
}i,j\mathbf{)}$\textbf{\ }stores the number of times that item $j$ has Borda
score $i$ for $i=0,...,d-1,$ see $\left[ 6,\ \text{p.17}\right] $. Table 2
displays the matrix \textbf{M} for the toy example $\mathbf{R}$ displayed in
Table 1. We note the following facts:

a) It has uniform row and column marginals equal to the sample size.

b) We can compute the Borda scale $\mathbf{\beta }$ from it.

c) It reveals the nature of crossing of scores attributed to the items for a
given binary partition of the items. For the toy example, consider the
partition $\left\{ C\right\} $ and $\left\{ B,A\right\} $ with attributed
scores of $\left\{ 0\right\} $ and $\left\{ 1,2\right\} $ respectively (this
is the first step in a riffle shuffle). Then the highlighted cells (marked
in bold) in Table 2 show that there are two crossing of scores, permutation
(transposition) of the scores 0 and 1, between the sets $\left\{ C\right\} $
and $\left\{ B,A\right\} $, (this is the second step in a random shuffle).
Furthermore the third row of Table 2 shows that the score 2 is equally
attributed to both items of the set $\left\{ B,A\right\} $ and it never
crossed to $\left\{ C\right\} $.

\begin{tabular}{|l|rrr|r|}
\multicolumn{5}{l}{\textbf{Table 2: The matrix of first-order marginals of
R.}} \\
 \hline
  & $C$ & $B$ & $A$ & row sum \\
 \hline
$0$ & 2 & \textbf{1} & \textbf{1} & 4 \\
$1$ & \textbf{2} & 1 & 1 & 4 \\
$2$ & 0 & 2 & 2 & 4 \\ \hline\hline
column sum & 4 & 4 & 4 &  \\ \hline
Borda scale\ $\mathbf{\beta }$ & 0.5 & 1.25 & 1.25 &  \\ \hline
\end{tabular}

\subsection{Coded tables $\mathbf{R}_{double}$ and $\mathbf{R}_{nega}$}

Our methodological approach is based on Benz\'{e}cri's platform, see $\left[
13,\ p.1113\right] ,$ that we quote: \textquotedblleft\ the main problem
inductive statistics has to face is to \textit{build tables} that, through
appropriate \textit{coding} and \textit{eventual supplementation}, give to
the available data such a shape that the analysis is able to extract from it
the answer to any \textit{question that we are allowed to ask}%
\textquotedblright . Italics are ours.

There are three elements in Benz\'{e}cri's platform: a) \textit{coding, }a
kind of pre-processing of data, will be discussed in the following
paragraph; b) \textit{eventual supplementation} consists in applying TCA and
not correspondence analysis (CA), because in the CA case we do not have a
result similar to Theorem 1; c) \textit{question that we are allowed to ask }%
is to explore and visualize rank data.

Within the CA framework, there are two codings of rank data $\mathbf{R}%
_{double}$ and $\mathbf{R}_{nega.}$.

\subsubsection{$\mathbf{R}_{double}$}

The first one is the doubled table of size $(2n)\times d$%
\begin{equation*}
\mathbf{R}_{double}=(_{\overline{\mathbf{R}}}^{\mathbf{R}})
\end{equation*}%
proposed independently by $\left[ 14,\ 15\right] $, where they showed that
CA of $\mathbf{R}_{double}$ is equivalent to the dual scaling of Nishisato
coding of rank data, see $\left[ 16\right] $. CA of $\mathbf{R}_{double}$ is
equivalent to CA of its first residual correspondence matrix%
\begin{equation*}
\mathbf{P}_{double}^{1}=\frac{1}{t}(_{-(r_{ij}-\frac{d-1}{2})}^{(r_{ij}-%
\frac{d-1}{2})}),
\end{equation*}%
where $t=nd(d-1)$. The structure of $\mathbf{P}_{double}^{1}$ shows that
each row is centered as in Carroll's multidimensional preference analysis
procedure, MDPREF, exposed in Alvo and Yu (2014, p.15). In TCA the objective
function to maximize is a combinatorial problem, see equation (3); and the
first iteration in TCA of $\mathbf{R}_{double}$ corresponds to computing
\begin{equation*}
\begin{array}{cccc}
\delta _{1}^{double} &=&\max_{\mathbf{v\in }\left\{ -1,1\right\} ^{n}}||(%
\mathbf{v}^{t}\ |\ \mathbf{-v}^{t})\mathbf{P}_{double}^{1}||_{1} & \tag{11}
\\
&=&\max_{\mathbf{v\in }\left\{ -1,1\right\} ^{n}}\frac{2}{t}%
\sum_{j=1}^{d}|\sum_{i=1}^{n}(r_{ij}-\frac{d-1}{2})v_{i}|\text{.} &  \notag
\end{array}
\end{equation*}

\subsubsection{$\mathbf{R}_{nega}$}

In the second approach, we summarize $\overline{\mathbf{R}}$ by its column
total; that is, we create a row named $\mathbf{nega=}$ $n\overline{\mathbf{%
\beta }}=\mathbf{1}_{n}^{\prime }\overline{\mathbf{R}},$ then we vertically
concatenate $\mathbf{nega}$ to $\mathbf{R}$, thus obtaining%
\begin{equation*}
\mathbf{R}_{nega}=(_{\mathbf{nega}}^{\mathbf{R}})
\end{equation*}%
of size $(n+1)\times d.$

$\left[ 17\right] $ discussed the relationship between TCA of $\mathbf{R}%
_{double}$ and TCA of $\mathbf{R}_{nega}$: TCA of $\mathbf{R}_{nega}$ can be
considered as constrained TCA of $\mathbf{R}_{double}$, because we are
constraining the vector $\mathbf{-v}^{t}=\mathbf{-1}_{n}^{t}$ in (11); that
is, the objective function to maximize corresponds to computing
\begin{equation}
\begin{array}{cccc}
\delta _{1} &=&\max_{\mathbf{v\in }\left\{ -1,1\right\} ^{n}}||(\mathbf{v}%
^{t}\ |\ \mathbf{-1}_{n}^{t})\mathbf{P}_{double}^{1}||_{1} & \tag{12} \\
&=&\max_{\mathbf{v\in }\left\{ -1,1\right\} ^{n}}||(\mathbf{v}^{t}\ \
\mathbf{-}1)\mathbf{P}_{nega}^{1}||_{1} \\
&=&\max_{\mathbf{v\in }\left\{ -1,1\right\} ^{n}}\frac{1}{t}%
\sum_{j=1}^{d}|\sum_{i=1}^{n}(r_{ij}-\frac{d-1}{2})(v_{i}+1)|\text{.} & \notag
\end{array}
\end{equation}%
So, we see that if in (11) the optimal value of $\mathbf{v}=\mathbf{1}_{n}$,
then $\delta _{1}^{double}=\delta _{1},$ otherwise $\delta
_{1}^{double}>\delta _{1}$.

Let
\begin{equation*}
\mathbf{v}_{1}=\arg \max_{\mathbf{v\in }\left\{ -1,1\right\} ^{n}}\frac{1}{t}%
\sum_{j=1}^{d}|\sum_{i=1}^{n}(r_{ij}-\frac{d-1}{2})(v_{i}+1)|.
\end{equation*}%
Define the set of indices $I_{+}=\left\{ i|v_{1i}=1\right\} $ and $%
I_{-}=\left\{ i|v_{1i}=-1\right\} ,$ where $\mathbf{v}_{1}=(v_{1i}).$ Then
\begin{equation}
\delta _{1}=\frac{2}{t}\sum_{j=1}^{d}|\sum_{i\in I_{+}}(r_{ij}-\frac{d-1}{2}%
)|  \tag{13}
\end{equation}%
shows that the summation in (13) is restricted to the subset of assessors
that belong to $I_{+}$. The subset $I_{+}$ indexes the voters having the
same direction in their votes. Given that we are uniquely interested in the
first TCA dimension, all the necessary information is encapsulated in $I_{+}$%
, as discussed in $\left[ 17,\ 9\right] $ using other arguments.

Furthermore, $\delta _{1}$ in (13) equals four times the cut norm of $%
\mathbf{R}_{centered}(i,j)=\frac{1}{t}(r_{ij}-\frac{d-1}{2}),$ where the cut
norm is defined to be
\begin{eqnarray*}
\left\vert \left\vert \mathbf{R}_{centered}\right\vert \right\vert _{cut}
&=&\max_{S,T}\frac{1}{t}\sum_{j\in S}\sum_{i\in T}(r_{ij}-\frac{d-1}{2}) \\
&=&\frac{1}{t}\sum_{j\in S_{+}}\sum_{i\in I_{+}}(r_{ij}-\frac{d-1}{2}) \\
&=&\delta _{1}/4,
\end{eqnarray*}%
where $S\subseteq \left\{ 1,...,d\right\} $ and $T\subseteq I;$ it shows
that the subsets $I_{+}$ and $S_{+}$ are positively associated, for further
details see for instance, $\left[ 18,\ 19\right] $.

In the sequel, we will consider only the application of TCA to $\mathbf{R}%
_{nega}$.

\section{First TCA voter factor scores of R$_{nega}$}

We show the results on the SUSHI data set enumerating $n=5000$ preferences
of $d=10$ sushis, see $\left[ 1\right] $. Even though, our interest concerns
only the first TCA voter factor scores of a voting profile $V_{1},$ it is a
common practice in CA circles to present the principal map of the row and
column projections.

Figures 1 and 2 display the principal maps obtained from CA and TCA of $%
R_{nega}$ of the SUSHI data denoted by $V_{1}$. We observe that, TCA
clusters the voters into a finite number of discrete patterns, while CA does
not: This is the main reason that we prefer the use of TCA to the use of the
classical well known dimension reduction technique CA.

We have the following theorem concerning the first TCA principal factor
scores of the voters belonging to a profile $V_{1}$, $f_{1}(i)$ for $%
i=1,...,n$, where the first principal axis partitions the $d$ items into $%
d_{1}$ and $d_{2}$ parts such that $d=d_{1}+d_{2}.$

\begin{figure}
     \centering
     \begin{subfigure}{2.0in}
         \centering
         \includegraphics[width=2.5in]{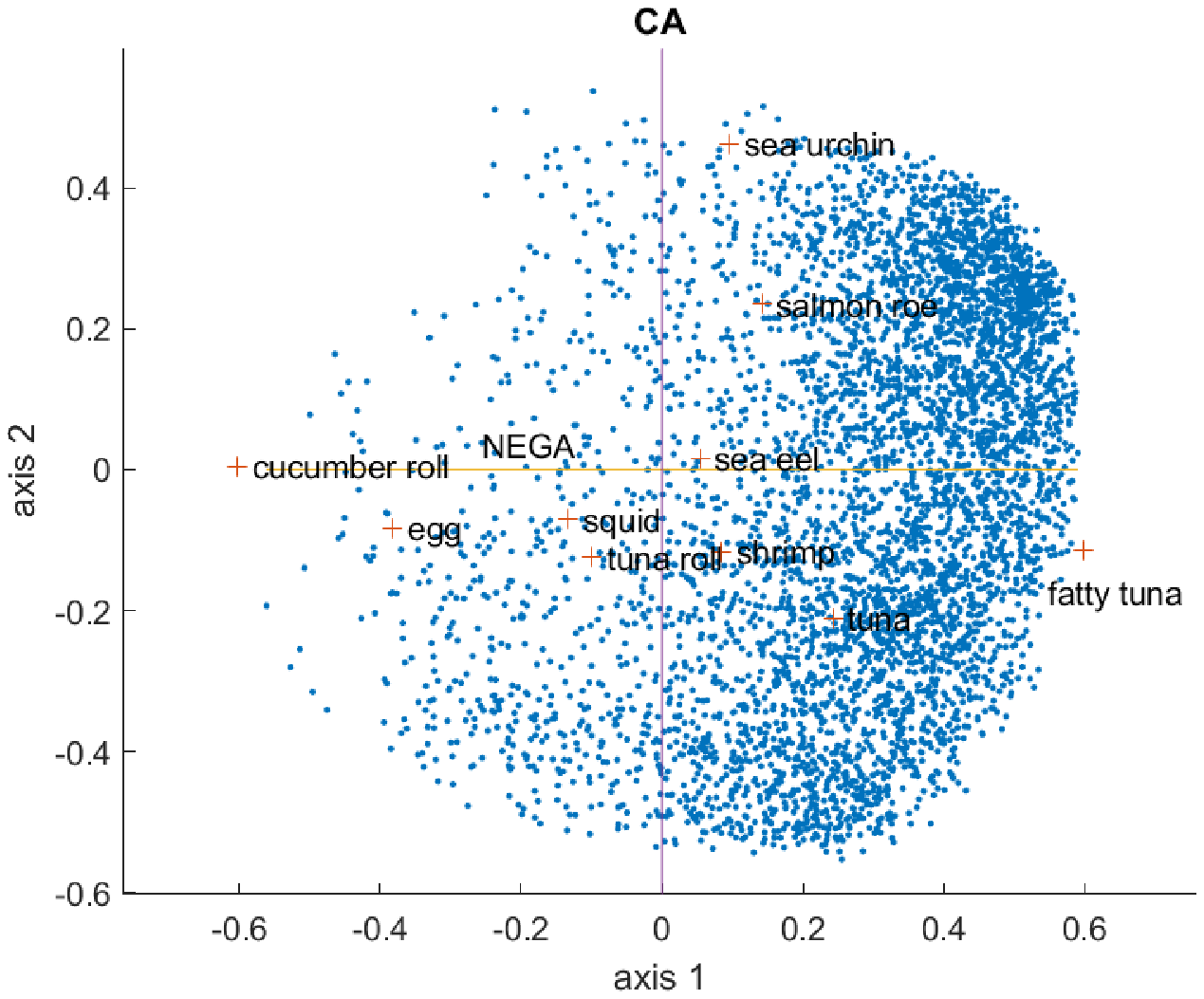}
         \caption{{\tiny Figure 1:CA map of SUSHI rank data}}
     \end{subfigure}
      \hspace{1.0in}
     \begin{subfigure}{2.0in}
         \centering
         \includegraphics[width=2.5in]{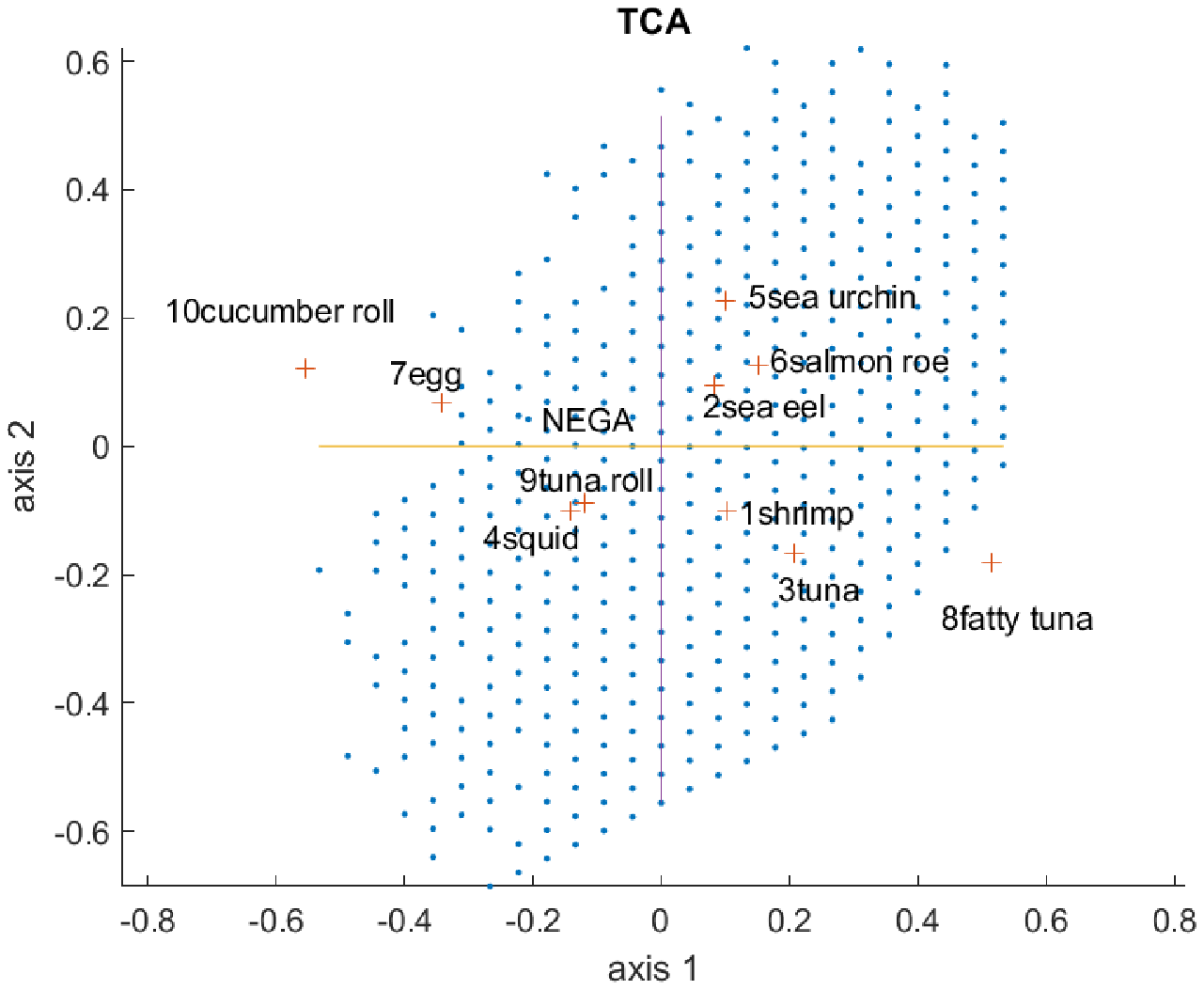}
         \caption{ {\tiny Figure 2: TCA map of SUSHI rank data} }
     \end{subfigure}

\end{figure}

\textbf{Theorem 1}

a) The maximum number of distinct clusters of the $n$ voters belonging to $%
V_{1}$ on the first TCA principal axis (distinct $f_{1}(i\mathbf{)}$ values
for $i\mathbf{\in }V_{1})$ is $d_{1}d_{2}+1.$

b) The maximum value that $f_{1}(i\mathbf{)}$ can attain is $2\frac{%
d_{1}d_{2}}{d(d-1)}.$

c) The minimum value that $f_{1}(i\mathbf{)}$ can attain is $-2\frac{%
d_{1}d_{2}}{d(d-1)}.$

d) If the number of distinct clusters is maximum, $d_{1}d_{2}+1$, then the
gap between two contiguous $f_{1}(i\mathbf{)}$ values is $\frac{4}{d(d-1)}%
.\bigskip $

\textbf{Remark 1}

a) We fix $f_{1}(nega)<0$ to eliminate the sign indeterminacy of the first
bilinear term in (10).

b) We partition $V_{1}$ into $d_{1}d_{2}+1$ clusters, $V_{1}=\cup _{\alpha
=1}^{d_{1}d_{2}+1}V_{1,\alpha }$, where the voters of the $\alpha $th
cluster are characterized by their first TCA factor score; that is, $%
V_{1,\alpha }=\left\{ i\in V_{1}\mathbf{:}f_{1}^{V_{1}}(i\mathbf{)=}2\frac{%
d_{1}d_{2}}{d(d-1)}-(\alpha -1)\frac{4}{d(d-1)}\right\} $ for $\alpha
=1,...,d_{1}d_{2}+1$.\bigskip

\textbf{Example 1:} In Figure 2, $d_{1}=4$ and $d_{2}=6,$ and we observe

Fact 1: by Theorem 1a, 5000 preferences are clustered into $d_{1}d_{2}+1=25$
clusters on the first TCA principal axis.

Fact 2: by Theorem 1b, the maximum value of $f_{1}(i\mathbf{)=}$ $48/90%
\mathbf{.}$

Fact 3: by Theorem 1c, the minimum value of $f_{1}(i\mathbf{)=}$ $-48/90%
\mathbf{.}$

Fact 4: by Theorem 1d, the gap separating two contiguous clusters of voters
on the first TCA principal axis is $4/90.\bigskip $

A cluster of voters defined in Remark 1b, $V_{1,\alpha }$ for $\alpha
=1,...,d_{1}d_{2}+1,$ can be classified as coherent or incoherent. And this
will be discussed in the next section.

\section{Coherent cluster}

The following definition characterizes a coherent cluster.\bigskip

\textbf{Definition 1 (}Coherency of a cluster of voters $V_{1,\alpha }$ for $%
\alpha =1,...,d_{1}d_{2}+1$\textbf{)}

A cluster of voters $V_{1,\alpha }$ $\subseteq V_{1}$ is coherent if $%
f_{1}^{V_{1,\alpha }}(v\mathbf{)=}2\frac{d_{1}d_{2}}{d(d-1)}-(\alpha -1)%
\frac{4}{d(d-1)}$ for all $v\mathbf{\in }V_{1,\alpha },$ where $%
f_{1}^{V_{1,\alpha }}(i\mathbf{)}$ is the first TCA factor score of the
voter $i\mathbf{\in }V_{1,\alpha }$ obtained from TCA of subprofile $%
V_{1,\alpha }.\bigskip $

\textbf{Remark 2:}

a) It is important to distinguish between $f_{1}^{V_{1}}(i\mathbf{)}$ for $%
i=1,...,|V_{1}|$ where $n=|V_{1}|,$ and $f_{1}^{V_{1,\alpha }}(i\mathbf{)}$
for $i=1,...,|V_{1,\alpha }|,$ where $|V_{1,\alpha }|$ represents the sample
size of the cluster $|V_{1,\alpha }|.$

b) Definition 1 implies that a cluster $V_{1,\alpha }$ is coherent when for
all voters $i\mathbf{\in }V_{1,\alpha }$ the first TCA factor score $%
f_{1}^{V_{1,\alpha }}(i\mathbf{)}$ does not depend on the voter $i$, but it
depends on $(\alpha ,d_{1},d_{2}).\bigskip $

\textbf{Corollary 1:} It follows from Remark 1a and equation (13) that, a
necessary condition, but not sufficient, for a cluster $V_{1,\alpha }$ to be
coherent is that its first TCA factor score obtained from TCA of $V_{1}$ is
strictly positive; that is, $0<f_{1}^{V_{1}}(i)$ for $i\in V_{1,\alpha }.$

\begin{figure}
\begin{subfigure}{.5\textwidth}
  \centering
  \includegraphics[width=.8\linewidth]{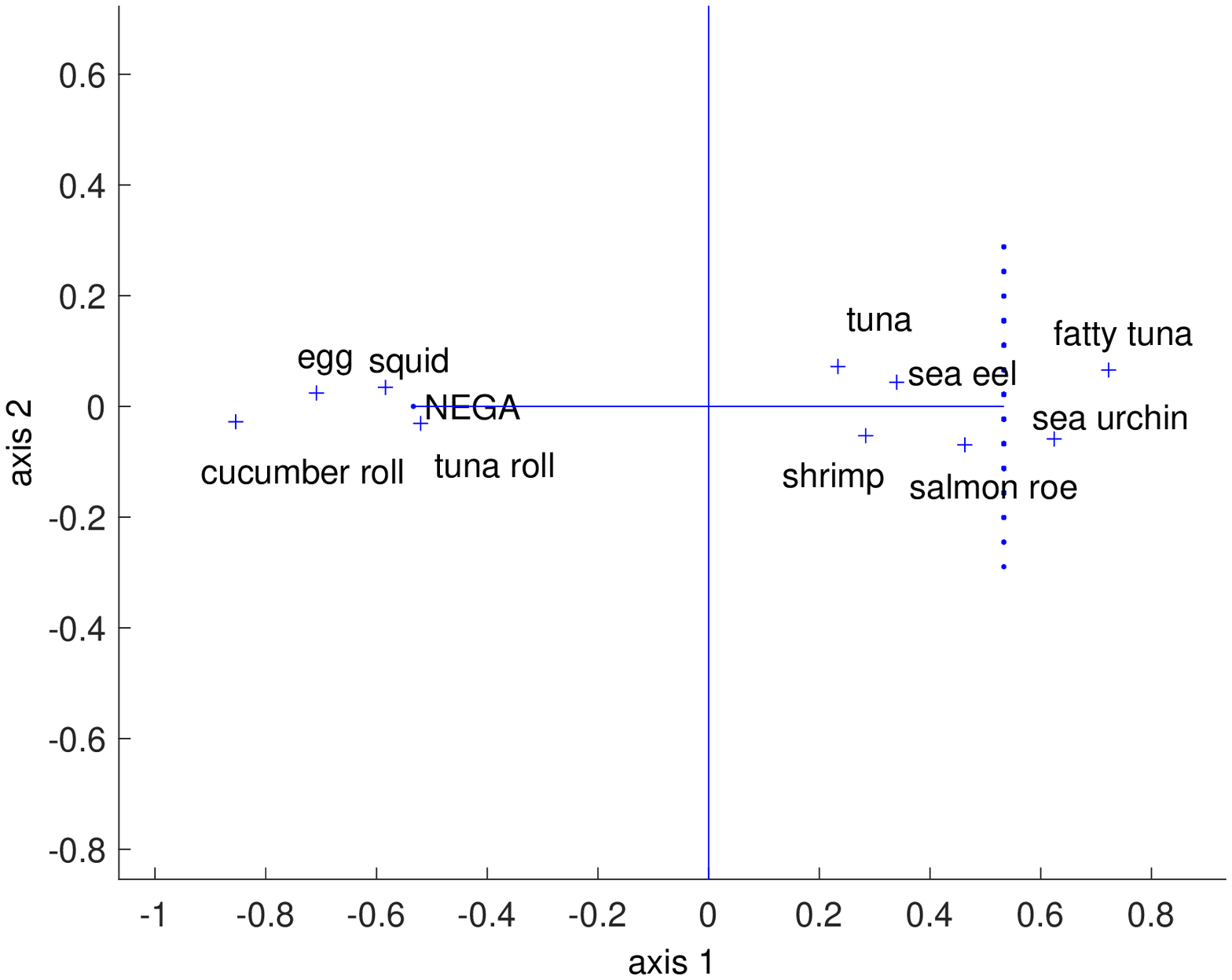}
  \caption{Figure 3 : TCA map of $\mbox{V}_{1,1}$.}
\end{subfigure}
\begin{subfigure}{.5\textwidth}
  \centering
  \includegraphics[width=.8\linewidth]{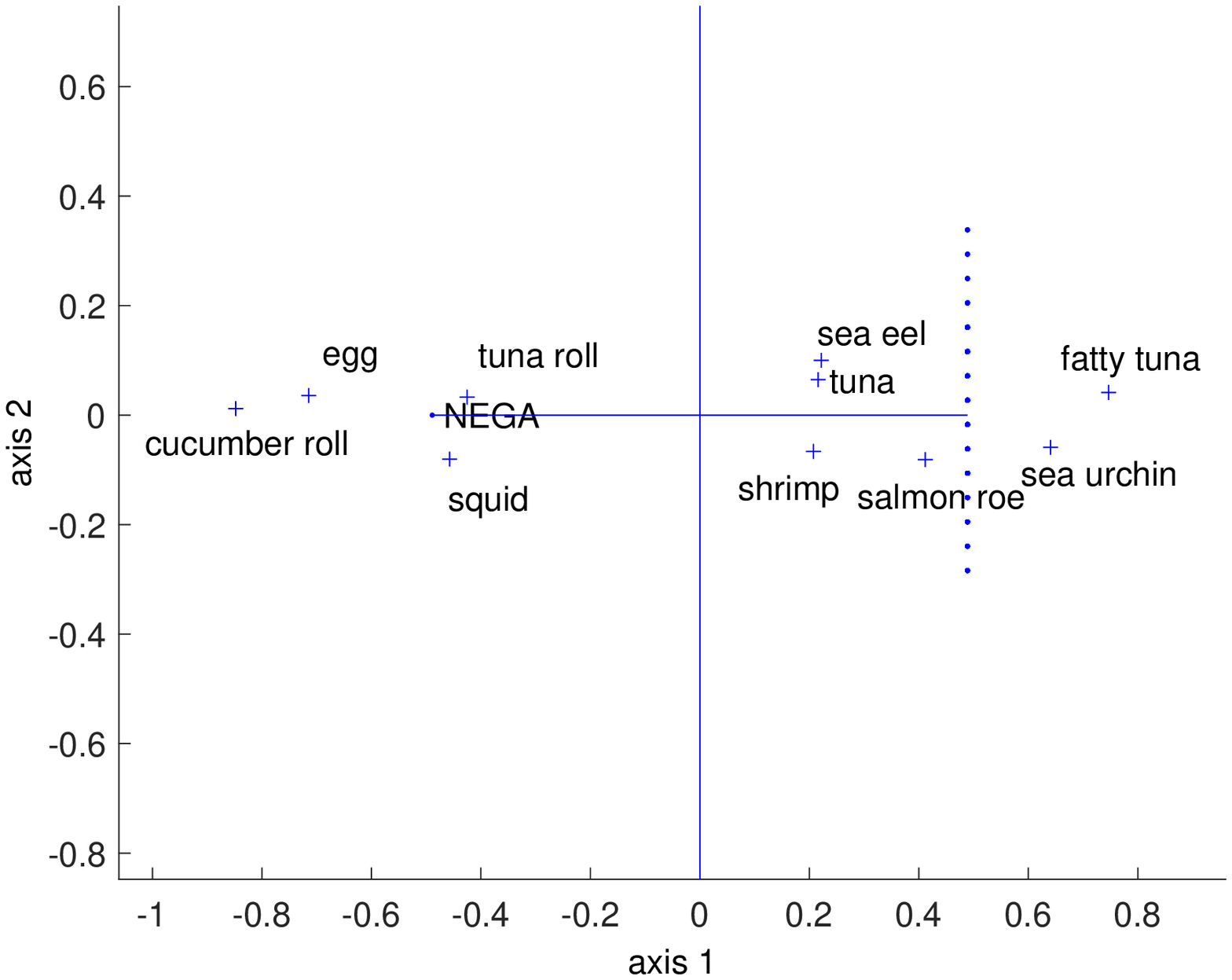}
  \caption{Figure 4 :  TCA map of $\mbox{V}_{1,2}$.}
\end{subfigure}

\begin{subfigure}{.5\textwidth}
  \centering
  \includegraphics[width=.8\linewidth]{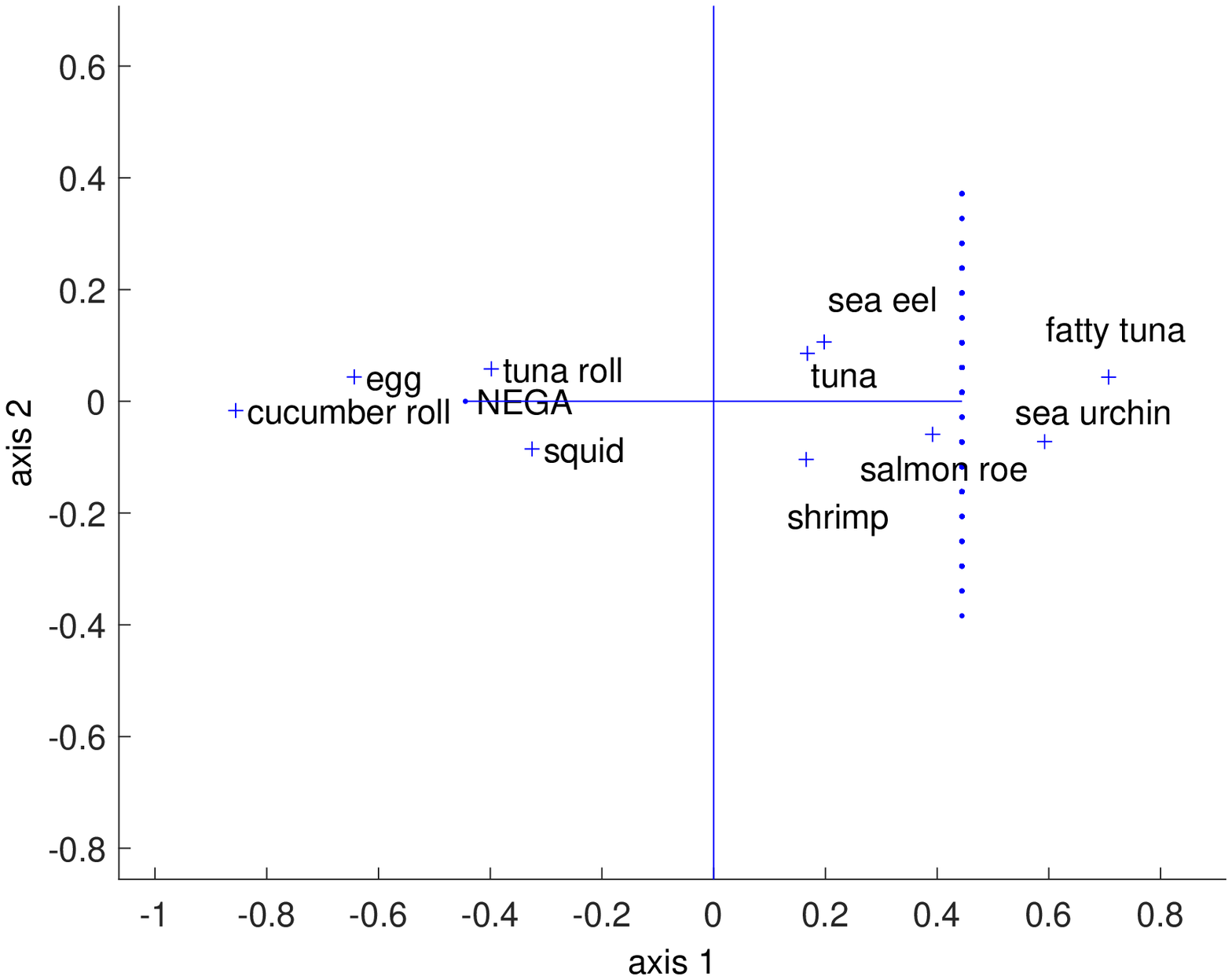}
  \caption{Figure 5 : TCA map of $\mbox{V}_{1,3}$.}
\end{subfigure}
\begin{subfigure}{.5\textwidth}
  \centering
  \includegraphics[width=.8\linewidth]{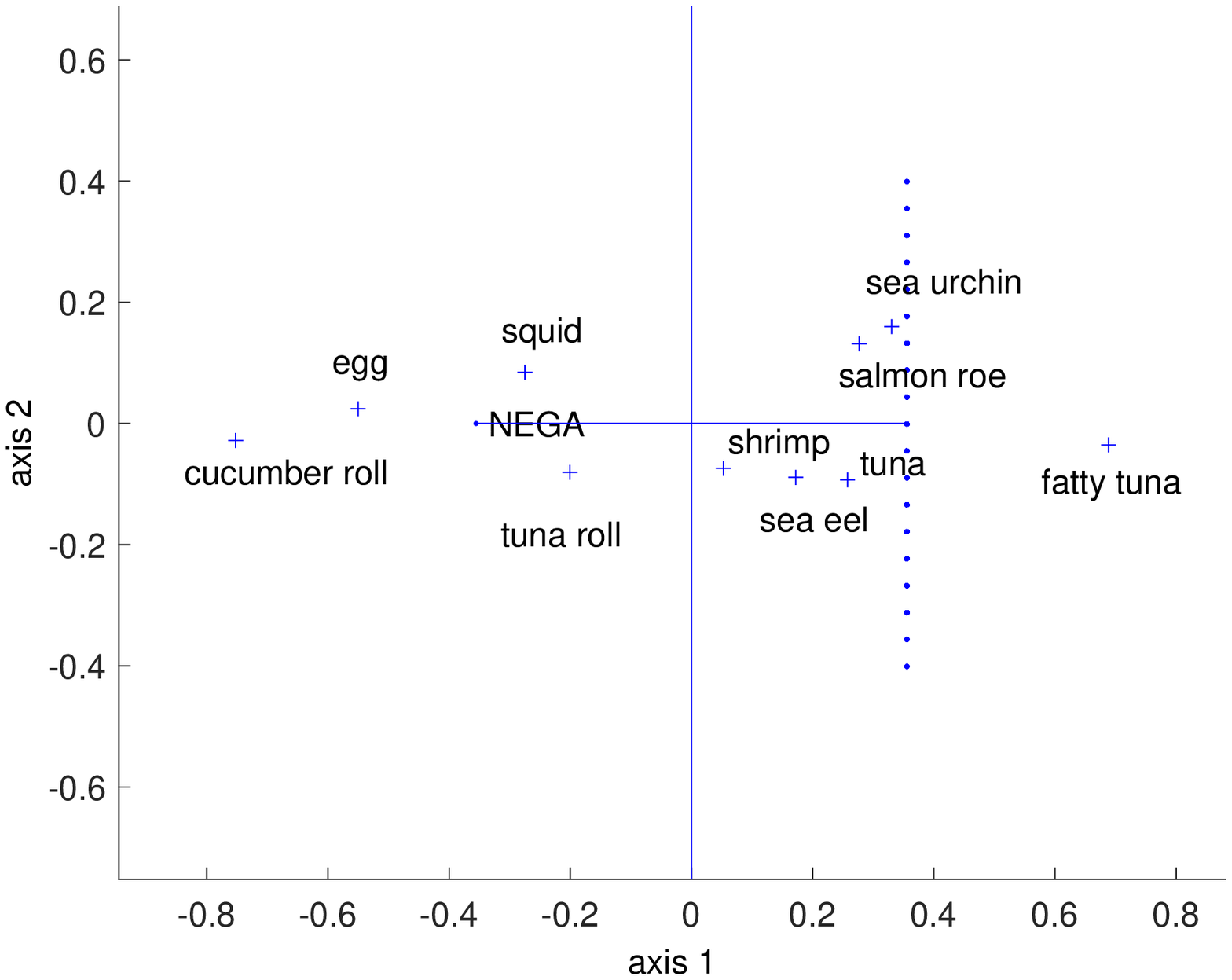}
  \caption{Figure 6 : TCA map of $\mbox{V}_{1,4}$.}
  \label{fig:sub-fourth}
\end{subfigure}


\begin{subfigure}{.5\textwidth}
  \centering
  \includegraphics[width=.8\linewidth]{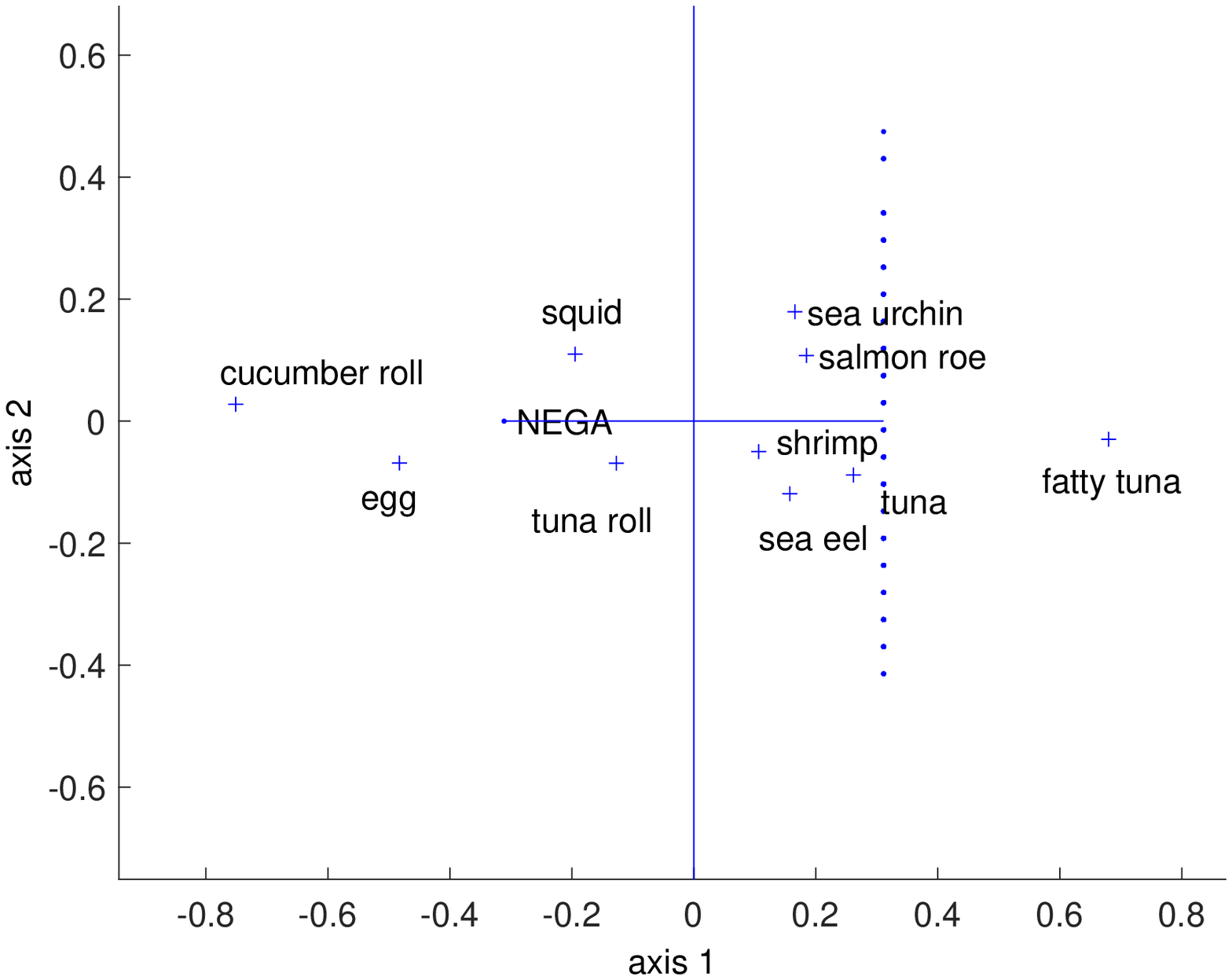}
  \caption{Figure 7 : TCA map of $\mbox{V}_{1,5}$.}
  \label{fig:sub-third}
\end{subfigure}
\begin{subfigure}{.5\textwidth}
  \centering
  \includegraphics[width=.8\linewidth]{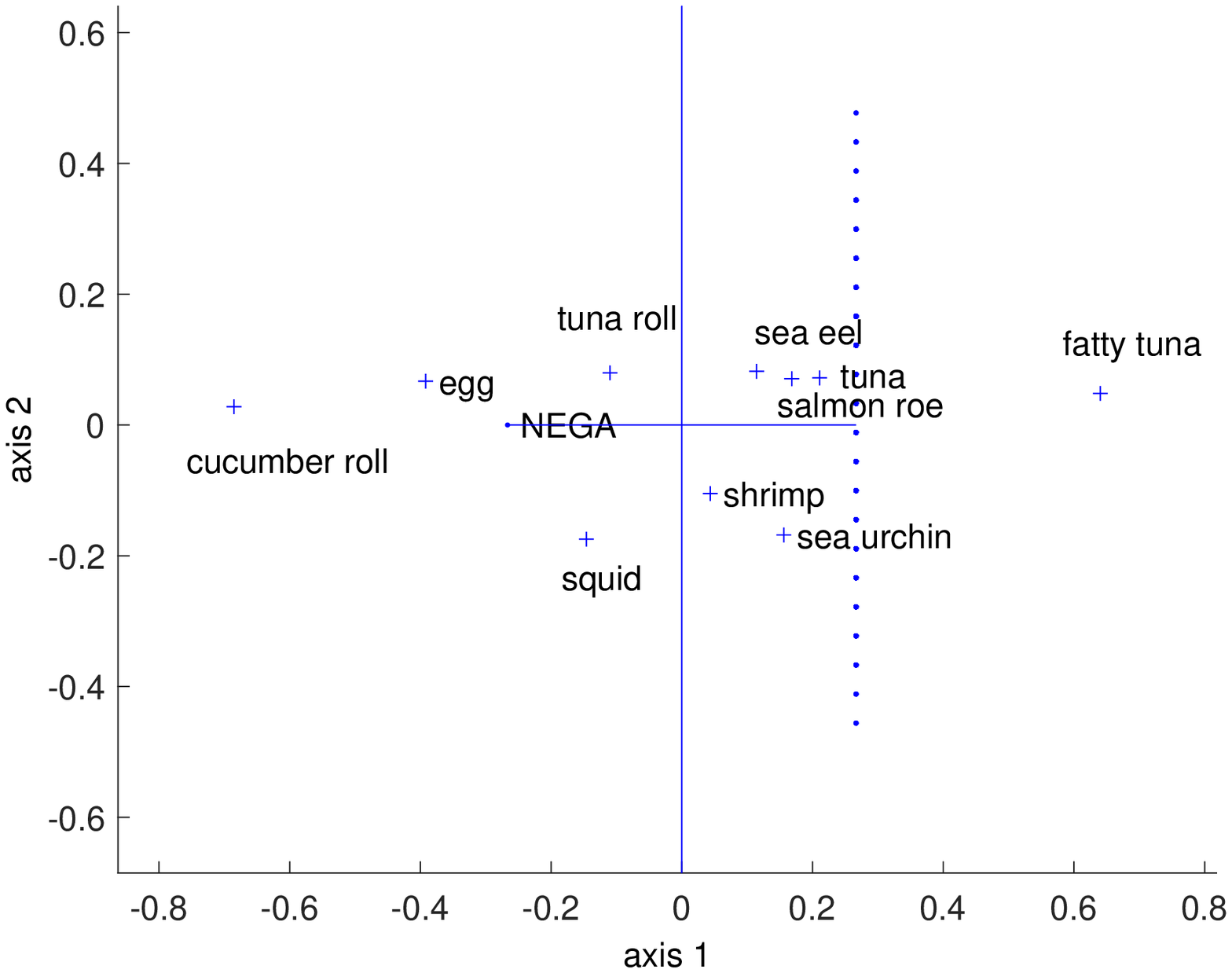}
  \caption{Figure 8 :  TCA map of $\mbox{V}_{1,6}$.}
  \label{fig:sub-fourth}
\end{subfigure}
\begin{subfigure}{.5\textwidth}
  \centering
  \includegraphics[width=.8\linewidth]{Fig9SUSHIG1C7ind401.eps}
  \caption{Figure 9 : TCA map of $\mbox{V}_{1,7}$.}
  \label{fig:sub-third}
\end{subfigure}
\begin{subfigure}{.5\textwidth}
  \centering
  \includegraphics[width=.8\linewidth]{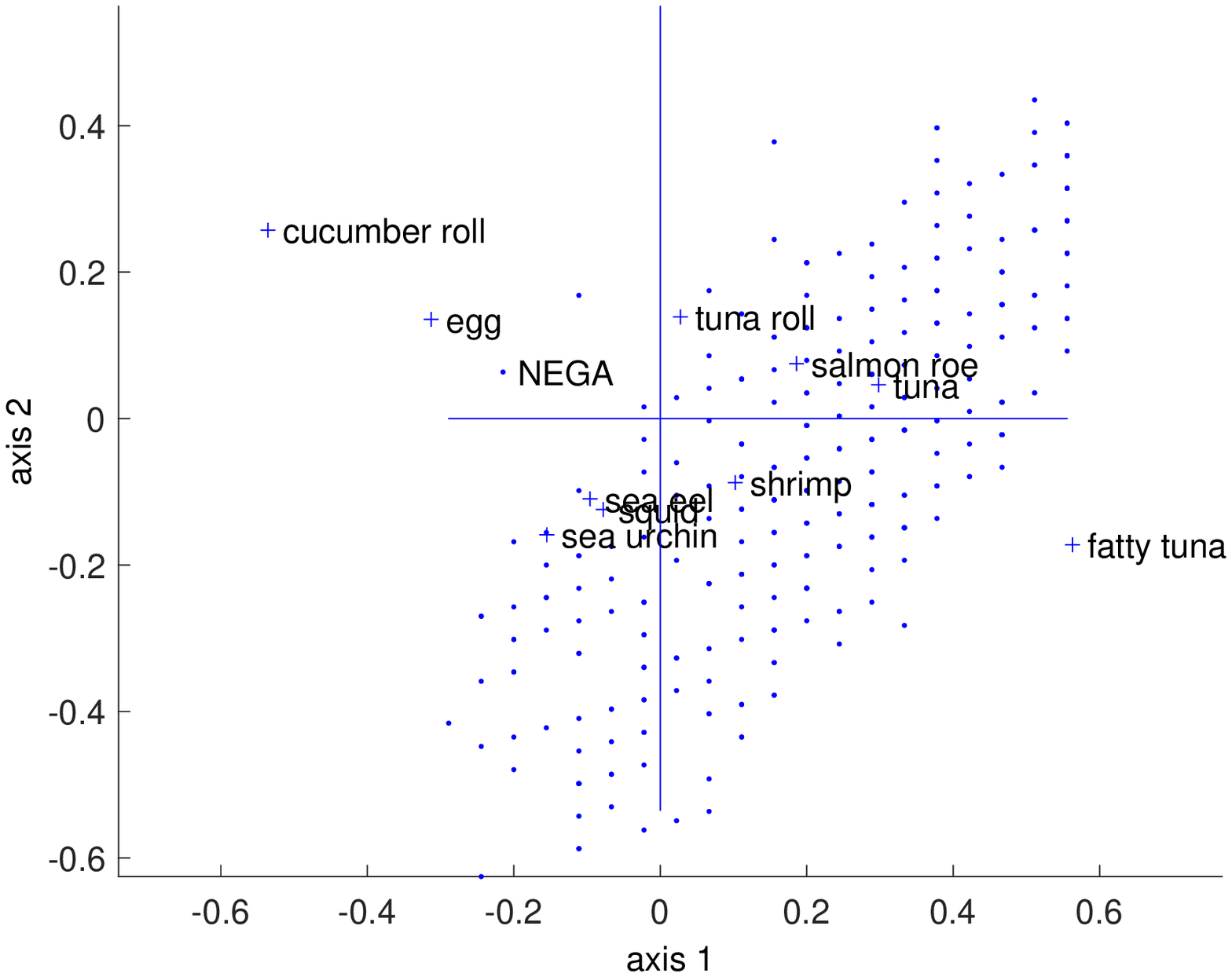}
  \caption{Figure 10:  TCA map of $\mbox{V}_{1,8}$.}
  \label{fig:sub-fourth}
\end{subfigure}
\end{figure}
\textbf{Example 2:} Figures 3 through 9 show the coherency of the clusters
of voters $V_{1,\alpha }$ for $\alpha =1,...,7,$ where dots represent
clusters of voters$;$ while Figure 10 shows the incoherence of the cluster $%
V_{1,8}.$ Further, the first three columns of Table 3 display the
mathematical formulation of the 7 coherent clusters $cohC_{1}(\alpha
)=V_{1,\alpha }$ for $\alpha =1,...,7$ as defined in Remark 1b and their
sample sizes $|V_{1,\alpha }|.$\bigskip

{\small\begin{tabular}{|l|l|l|l|l|l|}
\multicolumn{6}{l}{\textbf{Table 3: Characteristics of }$cohC_{1}(\alpha )=$
$V_{1,\alpha }$ \textbf{of SUSHI data}.} \\ \hline
$\alpha $ & $|V_{1,\alpha }|$ & description of $V_{1,\alpha }$ & $\delta
_{1}(V_{1,\alpha })$ & $T_{v\in V_{1,\alpha }}(\tau _{J_{1}}(S_{1}))$ & $%
Cross(V_{1,\alpha })$ \\ \hline
$1$ & $314$ & $\left\{ i\mathbf{:}f_{1}^{V_{1}}(i\mathbf{)}=48/90\right\} $
& $48/90$ & $6$ & $0$ \\
$2$ & $235$ & $\left\{ i\mathbf{:}f_{1}^{V_{1}}(i\mathbf{)}=44/90\right\} $
& $44/90$ & $7$ & $1/12$ \\
$3$ & $326$ & $\left\{ i\mathbf{:}f_{1}^{V_{1}}(i\mathbf{)}=40/90\right\} $
& $40/90$ & $8$ & $2/12$ \\
$4$ & $315$ & $\left\{ i\mathbf{:}f_{1}^{V_{1}}(i\mathbf{)}=36/90\right\} $
& $36/90$ & $9$ & $3/12$ \\
$5$ & $452$ & $\left\{ i\mathbf{:}f_{1}^{V_{1}}(i\mathbf{)}=32/90\right\} $
& $32/90$ & $10$ & $4/12$ \\
$6$ & $375$ & $\left\{ i\mathbf{:}f_{1}^{V_{1}}(i\mathbf{)}=28/90\right\} $
& $28/90$ & $11$ & $5/12$ \\
$7$ & $401$ & $\left\{ i\mathbf{:}f_{1}^{V_{1}}(i\mathbf{)}=24/90\right\} $
& $24/90$ & $12$ & $6/12$ \\ \hline
\end{tabular}
\bigskip

\textbf{Proposition 1:} For a voting profile $V$, $\delta _{1}(V)\geq
|f_{1}(nega)|$, where $\delta _{1}(V)$ is the\ first TCA dispersion value
obtained from TCA of $V,$ and $f_{1}(nega)$ is the first TCA factor score of
the row $nega$.\bigskip

The equality in Proposition 1 is attained only for coherent clusters as
shown in the following result.\bigskip

\textbf{Proposition 2:} The first TCA dispersion value of a coherent cluster
$cohC_{1}(\alpha )$ satisfies
\begin{eqnarray*}
\delta _{1}(cohC_{1}(\alpha )) &=&|f_{1}^{V_{1,\alpha }}(nega)|. \\
&\mathbf{=}&2\frac{d_{1}d_{2}}{d(d-1)}-(\alpha -1)\frac{4}{d(d-1)}
\end{eqnarray*}%
\bigskip

\textbf{Example 3:}  propostion 2 can be observed by looking at the columns
3 and 4 of Table 3 which concern the 7 coherent clusters $cohC_{1}(\alpha
)=V_{1,\alpha }$ for $\alpha =1,...,7$. While for the incoherent cluster $%
V_{1,8}$ with sample size of $|V_{1,8}|=335,$ we observe: $V_{1,8}=\left\{ i%
\mathbf{:}f_{1}^{V_{1}}(i\mathbf{)}=20/90=0.222\right\} ,$ and by
Proposition 1, $\delta _{1}(V_{1,8})=0.2354$ $>2/9.$ This means that the 335
voters belonging to $V_{1,8}$ form a cluster within the whole sample of 5000
voters, but separated as 335 voters they do not form a coherent cluster.

\subsection{Interpretability of a coherent cluster}

The following result shows that for coherent clusters, the first TCA
dimension can be interpreted as Borda scaled factor.\bigskip

\textbf{Proposition 3:} The first TCA column factor score of the item $j,$ $%
g_{1}(j),$ is an affine function of the Borda scale $\beta (j);$ that is, $%
g_{1}(j)=\frac{2}{d-1}\beta (j)-1$\ for $j=1,...,d.$\ Or\ $corr(\mathbf{g}%
_{1},\mathbf{\beta })=1.\bigskip $

\textbf{Remark 3:}

The first TCA principal factor score of item $j$ for $j=1,...,d$ is bounded:
$-1\leq g_{1}(j)\leq 1,$\ because $0\leq \beta (j)\leq d-1.\bigskip $

\textbf{Example 4:} Table 4 displays the Borda scales of the items, sushis,
in the seven coherent clusters $cohC_{1}(\alpha )=V_{1,\alpha }$ for $\alpha
=1,...,7.$ To identify the sushi type, one has to refer to Figure 2; for
instance, $j10$ corresponds to $10cucumber$\ $roll$ in Figure 2. We observe
the following main fact: For each of the seven coherent clusters, the first
TCA principal axis produced the same binary partition of the items: $%
J_{1}=\left\{ j10,j7,j4,j9\right\} $ characterized by $4.5>\beta (j_{1})$
for $j_{1}\in J_{1}$, and $J_{2}=\left\{ j3,j1,j2,j6,j5,j8\right\} $
characterized by $\beta (j_{1})$ $>4.5$ for $j_{2}\in J_{2}.$ The six sushis
in $J_{2}$ have Borda scales above average score of $4.5=(0+9)/2$; while the
four sushis in $J_{1}$ have Borda scales below average score of $%
4.5.\bigskip \ $

{\tiny\begin{tabular}{|l|lrrr|rrrrrr|}
\multicolumn{11}{l}{\textbf{Table 4: Borda scales of the 10 sushis in the
seven coherent clusters.}} \\ \hline
Borda scale & \multicolumn{10}{|c|}{items} \\ \cline{2-11}
& \textbf{j10} & \textbf{j7} & \textbf{j4} & \textbf{j9} & j3 & j1 & j2 & j6
& j5 & j8 \\ \hline
$\mathbf{\beta }(cohC_{1}(1))$ & \multicolumn{1}{|c}{\textbf{0.66}} &
\multicolumn{1}{l}{\textbf{1.31}} & \textbf{1.87} & \textbf{2.16} & 5.55 &
\multicolumn{1}{l}{5.78} & 6.03 & 6.58 & 7.31 & 7.52 \\
$\mathbf{\beta }(cohC_{1}(2))$ & \multicolumn{1}{|c}{\textbf{0.69}} &
\multicolumn{1}{l}{\textbf{1.29}} & \textbf{2.44} & \textbf{2.59} & 5.47 &
\multicolumn{1}{l}{5.43} & 5.50 & 6.35 & 7.38 & 7.86 \\
$\mathbf{\beta }(cohC_{1}(3))$ & \multicolumn{1}{|c}{\textbf{0.65}} &
\multicolumn{1}{l}{\textbf{1.60}} & \textbf{3.04} & \textbf{2.71} & 5.25 &
\multicolumn{1}{l}{5.25} & 5.39 & 6.26 & 7.17 & 7.68 \\
$\mathbf{\beta }(cohC_{1}(4))$ & \multicolumn{1}{|c}{\textbf{0.83}} &
\multicolumn{1}{l}{\textbf{1.79}} & \textbf{3.10} & \textbf{3.28} & 5.30 &
\multicolumn{1}{l}{4.74} & 5.22 & 6.34 & 6.76 & 7.64 \\
$\mathbf{\beta }(cohC_{1}(5))$ & \multicolumn{1}{|c}{\textbf{1.12}} &
\multicolumn{1}{l}{\textbf{2.02}} & \textbf{3.26} & \textbf{3.60} & 5.70 &
\multicolumn{1}{l}{4.74} & 5.27 & 5.75 & 5.99 & 7.60 \\
$\mathbf{\beta }(cohC_{1}(6))$ & \multicolumn{1}{|c}{\textbf{1.12}} &
\multicolumn{1}{l}{\textbf{2.33}} & \textbf{3.62} & \textbf{3.93} & 5.68 &
\multicolumn{1}{l}{4.98} & 5.21 & 5.33 & 5.25 & 7.56 \\
$\mathbf{\beta }(cohC_{1}(7))$ & \multicolumn{1}{|c}{\textbf{1.42}} &
\multicolumn{1}{l}{\textbf{2.74}} & \textbf{3.84} & \textbf{4.00} & 5.45 &
\multicolumn{1}{l}{4.70} & 5.02 & 5.26 & 5.20 & 7.38 \\ \hline
\end{tabular}}
\bigskip

Now we ask the question what are the differences among the seven coherent
clusters? The answer is riffle shuffling of the scores of the items, which
we discuss next.

\section{Exploratory riffle shuffling}

$\left[ 8\right] $ is the seminal reference on riffle shuffling of cards. $%
\left[ 2\right] $ generalized the notion of independence of two subsets of
items to riffled independence to uncover the structure of rank data. Within
the framework of data analysis of preferences, exploratory riffle shuffling
can be described in the following way. We have two sets: $J$ a set of $d$
distinct items and $S$ a set of $d$ Borda scores. We partition both sets
into two disjoint subsets of sizes $d_{1}$ and $d_{2}=d-d_{1};$ that is, $%
J=J_{1}\cup J_{2}$ with $J_{1}=\left\{ j_{1},j_{2},...,j_{d_{1}}\right\} $
and $S=S_{1}\cup S_{2}$ with $S_{1}=\left\{ 0,1,...,d_{1}-1\right\} .$
Riffle shuffling consists of two steps. In the first step, we attribute the
scores of $S_{1}$ to $J_{1}$ and the scores of $S_{2}$ to $J_{2}.$ In the
second step, we permute some scores attributed to $J_{1}$ with the same
number of scores attributed to $J_{2}.$ The second step can be
mathematically described as an application of a permutation $\tau $, such
that $\tau _{J}(S_{1},S_{2})=(\tau _{J_{1}}(S_{1}),\tau _{J_{2}}(S_{2})).$
We interpret $\tau _{J_{1}}(S_{1})$ as the set of scores attributed to $%
J_{1},$ and $\tau _{J_{2}}(S_{2})$ as the set of scores attributed to $J_{2}.
$

\textbf{Example 5:} Table 5 displays a toy example with $n=7$ voters' Borda
scorings of $d=10$ items with $J_{1}=\left\{ a,b,c,d\right\} $ and $%
J_{2}=\left\{ e,f,g,h,i,j\right\} .$ We observe the following: a) The first
four voters have only done the first step in a riffle shuffle: each one of
them has attributed the scores in $S_{1}=\left\{ 0,1,2,3\right\} $ to the
items in $J_{1}$ and the scores in $S_{2}=\left\{ 4,5,6,7,8,9\right\} $ to
the items in $J_{2}.$ This can be described as $\tau
_{J}(S_{1},S_{2})=(S_{1},S_{2});$ that is the permutation $\tau $ is the
identity permutation; so there is no crossing of scores between $J_{1}$ and $%
J_{2}$. b) Voters 5, 6 and 7 have done both steps in a riffle shuffle.
Voters 5 and 6 have permuted score \textbf{3} with \textbf{5}, so we have $%
\tau _{J_{1}}(S_{1})=\left\{ 0,1,2,\mathbf{5}\right\} $ and $\tau
_{J_{2}}(S_{2})=\left\{ 4,\mathbf{3},6,7\mathbf{,}8,9\right\} $. Voter 7 has
permuted the scores $\left\{ \mathbf{2,3}\right\} $ with $\left\{ \mathbf{4,5%
}\right\} $, so we have $\tau _{J_{1}}(S_{1})=\left\{ 0,1,\mathbf{4},\mathbf{%
5}\right\} $ and $\tau _{J_{2}}(S_{2})=\left\{ \mathbf{2},\mathbf{3},6,7%
\mathbf{,}8,9\right\} $.

Further, we note by $|\tau _{J_{1}}(S_{1})|$ the number of voters who have
done the riffle shuffle $(\tau _{J_{1}}(S_{1}),\tau _{J_{2}}(S_{2}))$. So $%
|\tau _{J_{1}}(S_{1})=\left\{ 0,1,2,3\right\} |=4,$ $|\left\{ 0,1,2,\mathbf{5%
}\right\} |=2$ and $|\left\{ 0,1,\mathbf{4},\mathbf{5}\right\} |=1.$ The
permuted scores between the two blocks of items is in bold in Table 5.

\begin{tabular}{|l|cccc|cccccl|}
\multicolumn{11}{l}{\textbf{Table 5: Borda scorings of 10 items by 7 voters.}
} \\ \hline
voter & \multicolumn{10}{|c|}{items} \\ \cline{2-11}
& a & b & c & d & e & f & g & h & i & j \\ \hline
1 & 0 & 1 & 2 & 3 & 4 & 5 & 6 & 7 & 8 & 9 \\
2 & 0 & 2 & 3 & 1 & 6 & 4 & 5 & 8 & 7 & 9 \\
3 & 3 & 2 & 1 & 0 & 5 & 6 & 4 & 9 & 7 & 8 \\
4 & 2 & 1 & 0 & 3 & 8 & 7 & 9 & 4 & 5 & 6 \\
5 & 0 & 1 & 2 & \textbf{5} & 4 & \textbf{3} & 6 & 7 & 8 & 9 \\
6 & 1 & 2 & \textbf{5} & 0 & \textbf{3} & 6 & 4 & 9 & 7 & 8 \\
7 & 0 & \textbf{4} & \textbf{5} & 1 & 6 & 8 & 9 & \textbf{2} & 7 & \textbf{3}
\\ \hline
\end{tabular}%
\bigskip

\textbf{Remark 4:} A useful observation that we get from Example 5 is that
we can concentrate our study either on $J_{1}$ or on $J_{2}:$ For if we know
$\tau _{J_{1}}(S_{1})$, the scores attributed to $J_{1},$ we can deduce $%
\tau _{J_{2}}(S_{2})$, the scores attributed to $J_{2}$ because of mutual
exclusivity constraints ensuring that any two items, say $a$ and $b,$ never
map to the same rank by a voter. \bigskip

A simple measure of magnitude of $(d_{1},d_{2})$ riffle shuffling of a voter
$i$ is the sum of its Borda scores attributed to the items in $J_{1};$ that
is,
\begin{equation*}
T_{i}(\tau _{J_{1}}(S_{1}))=\sum_{j\in J_{1}}r_{ij},
\end{equation*}%
where $r_{ij}$ is the Borda score attributed to item $j$ by voter $i$. In
Table 5, for the first four voters, $T_{i}(\tau _{J_{1}}(S_{1}))=6$ for $%
i=1,...,4,$ which is the minimum attainable sum of scores; it implies that
for these voters there is no crossing of scores between the two blocks $%
J_{1} $ and $J_{2}$. While for voters 5 and 6, $T_{i}(\tau
_{J_{1}}(S_{1}))=8 $ for $i=5,6;$ for voter 7, $T_{7}(\tau
_{J_{1}}(S_{1}))=10$. These values show that the crossing of scores between
the two blocks $J_{1}$ and $J_{2}$ of voters 5 and 6 are at a lower level
than the crossing of scores for voter 7.

For relatively small sample sizes, it is easy to enumerate the different
types of $(d_{1},d_{2})$ riffle shuffles. For relatively large sample sizes,
we use the contingency table of first-order marginals, that we discuss next.

\subsection{Types of $(d_{1},d_{2})$ riffle shufflings in a coherent cluster}

The contingency table of first order marginals of an observed voting profile
$V$ on $d$ items is a square $d\times d$ matrix \textbf{M}, where $\mathbf{M(%
}i,j\mathbf{)}$\textbf{\ }stores the number of times that item $j$ has Borda
score $i$ for $i=0,...,d-1,$ see subsection 3.2. It helps us to observe
types of $(d_{1},d_{2})$ riffle shufflings in a coherent cluster as we
explain in Example 6.\bigskip

{\tiny\begin{tabular}{|l|lrrrrrrrrr|r|}
\multicolumn{12}{l}{\textbf{Table 6: M}$_{1,1},$ \textbf{contingency table
of first-order marginals of }$cohC_{1}(1).$} \\ \hline
Borda & \multicolumn{11}{|c|}{items} \\ \cline{2-12}
scores & j10 & j7 & j4 & j9 & j3 & j1 & j2 & j6 & j5 & j8 & sum \\ \hline
0 & \multicolumn{1}{||r}{\textbf{174}} & \textbf{92} & \textbf{37} & \textbf{%
11} & 0 & 0 & 0 & 0 & 0 & 0 & \multicolumn{1}{||r|}{314} \\
1 & \multicolumn{1}{||r}{\textbf{88}} & \textbf{88} & \textbf{76} & \textbf{%
62} & 0 & 0 & 0 & 0 & 0 & 0 & \multicolumn{1}{||r|}{314} \\
2 & \multicolumn{1}{||r}{\textbf{38}} & \textbf{78} & \textbf{91} & \textbf{%
107} & 0 & 0 & 0 & 0 & 0 & 0 & \multicolumn{1}{||r|}{314} \\
3 & \multicolumn{1}{||r}{\textbf{14}} & \textbf{56} & \textbf{110} & \textbf{%
134} & 0 & 0 & 0 & 0 & 0 & 0 & \multicolumn{1}{||r|}{314} \\
4 & \multicolumn{1}{||r}{0} & 0 & 0 & 0 & \textbf{92} & \textbf{78} &
\textbf{73} & \textbf{38} & \textbf{21} & \textbf{12} &
\multicolumn{1}{||r|}{314} \\
5 & \multicolumn{1}{||r}{0} & 0 & 0 & 0 & \textbf{95} & \textbf{77} &
\textbf{59} & \textbf{42} & \textbf{23} & \textbf{18} &
\multicolumn{1}{||r|}{314} \\
6 & \multicolumn{1}{||r}{0} & 0 & 0 & 0 & \textbf{47} & \textbf{63} &
\textbf{70} & \textbf{65} & \textbf{37} & \textbf{32} &
\multicolumn{1}{||r|}{314} \\
7 & \multicolumn{1}{||r}{0} & 0 & 0 & 0 & \textbf{35} & \textbf{49} &
\textbf{45} & \textbf{72} & \textbf{68} & \textbf{45} &
\multicolumn{1}{||r|}{314} \\
8 & \multicolumn{1}{||r}{0} & 0 & 0 & 0 & \textbf{32} & \textbf{27} &
\textbf{32} & \textbf{62} & \textbf{87} & \textbf{74} &
\multicolumn{1}{||r|}{314} \\
9 & \multicolumn{1}{||r}{0} & 0 & 0 & 0 & \textbf{13} & \textbf{20} &
\textbf{35} & \textbf{35} & \textbf{78} & \textbf{133} &
\multicolumn{1}{||r|}{314} \\ \cline{1-11}\cline{2-12}
$\beta $ & \multicolumn{1}{|c}{\textbf{0.66}} & \multicolumn{1}{|l|}{\textbf{%
1.31}} & \multicolumn{1}{|r}{\textbf{1.87}} & \multicolumn{1}{|r}{\textbf{%
2.16}} & \multicolumn{1}{|r}{5.55} & \multicolumn{1}{|l|}{5.78} &
\multicolumn{1}{|r}{6.03} & \multicolumn{1}{|r}{6.58} & \multicolumn{1}{|r}{
7.31} & \multicolumn{1}{|r|}{7.52} &  \\ \hline
\end{tabular}}

{\tiny\begin{tabular}{|l|lrrrrrrrrr|r|}
\multicolumn{12}{l}{\textbf{Table 7:\ M}$_{1,2},$ c\textbf{ontingency table
of first-order marginals of }$cohC_{1}(2).$} \\ \hline
\multicolumn{1}{|l|}{Borda} & \multicolumn{11}{|c|}{items} \\ \cline{2-12}
\multicolumn{1}{|l|}{scores} & j10 & j7 & j4 & j9 & j3 & j1 & j2 & j6 & j5 &
\multicolumn{1}{r|}{j8} & sum \\ \hline
0 & \multicolumn{1}{||r}{\textbf{127}} & \textbf{70} & \textbf{32} & \textbf{%
6} & 0 & 0 & 0 & 0 & 0 & 0 & \multicolumn{1}{||r|}{235} \\
1 & \multicolumn{1}{||r}{\textbf{69}} & \textbf{82} & \textbf{38} & \textbf{%
46} & 0 & 0 & 0 & 0 & 0 & 0 & \multicolumn{1}{||r|}{235} \\
2 & \multicolumn{1}{||r}{\textbf{32}} & \textbf{56} & \textbf{62} & \textbf{%
85} & 0 & 0 & 0 & 0 & 0 & 0 & \multicolumn{1}{||r|}{235} \\
3 & \multicolumn{1}{||r}{0} & 0 & 0 & 0 & \textit{55} & \textit{59} &
\textit{74} & \textit{29} & \textit{15} & \textit{3} & \multicolumn{1}{||r|}{
235} \\
4 & \multicolumn{1}{||r}{\textit{7}} & \textit{27} & \textit{103} & \textit{%
98} & 0 & 0 & 0 & 0 & 0 & 0 & \multicolumn{1}{||r|}{235} \\
5 & \multicolumn{1}{||r}{0} & 0 & 0 & 0 & \textbf{68} & \textbf{60} &
\textbf{42} & \textbf{41} & \textbf{11} & \textbf{13} &
\multicolumn{1}{||r|}{235} \\
6 & \multicolumn{1}{||r}{0} & 0 & 0 & 0 & \textbf{49} & \textbf{53} &
\textbf{35} & \textbf{48} & \textbf{32} & \textbf{18} &
\multicolumn{1}{||r|}{235} \\
7 & \multicolumn{1}{||r}{0} & 0 & 0 & 0 & \textbf{26} & \textbf{35} &
\textbf{42} & \textbf{48} & \textbf{40} & \textbf{44} &
\multicolumn{1}{||r|}{235} \\
8 & \multicolumn{1}{||r}{0} & 0 & 0 & 0 & \textbf{28} & \textbf{15} &
\textbf{22} & \textbf{44} & \textbf{70} & \textbf{56} &
\multicolumn{1}{||r|}{235} \\
9 & \multicolumn{1}{||r}{0} & 0 & 0 & 0 & \textbf{9} & \textbf{13} & \textbf{%
20} & \textbf{25} & \textbf{67} & \textbf{101} & \multicolumn{1}{||r|}{235}
\\ \cline{1-11}\cline{2-12}
$\beta $ & \multicolumn{1}{|c}{\textbf{0.69}} & \multicolumn{1}{|l|}{\textbf{%
1.29}} & \multicolumn{1}{|r}{\textbf{2.44}} & \multicolumn{1}{|r}{\textbf{%
2.59}} & \multicolumn{1}{|r}{5.47} & \multicolumn{1}{|l|}{5.43} &
\multicolumn{1}{|r}{5.50} & \multicolumn{1}{|r}{6.35} & \multicolumn{1}{|r}{
7.38} & \multicolumn{1}{|r|}{7.86} &  \\ \hline
\end{tabular}}

\bigskip
{\tiny\begin{tabular}{|l|lrrrrrrrrr|r|}
\multicolumn{12}{l}{\textbf{Table 8:\ M}$_{1,3},$ c\textbf{ontingency table
of first-order marginals of }$cohC_{1}(3).$} \\ \hline
\multicolumn{1}{|l|}{Borda} & \multicolumn{11}{|c|}{items} \\ \cline{2-12}
\multicolumn{1}{|l|}{scores} & j10 & j7 & j4 & j9 & j3 & j1 & j2 & j6 & j5 &
j8 & sum \\ \hline
0 & \multicolumn{1}{||r}{\textbf{182}} & \textbf{97} & \textbf{33} & \textbf{%
14} & 0 & 0 & 0 & 0 & 0 & 0 & \multicolumn{1}{||r|}{326} \\
1 & \multicolumn{1}{||r}{\textbf{104}} & \textbf{100} & \textbf{46} &
\textbf{76} & 0 & 0 & 0 & 0 & 0 & 0 & \multicolumn{1}{||r|}{326} \\
2 & \multicolumn{1}{||r}{\textit{19}} & \textit{41} & \textit{41} & \textit{%
70} & \textit{40} & \textit{37} & \textit{46} & \textit{17} & \textit{12} &
\textit{3} & \multicolumn{1}{||r|}{326} \\
3 & \multicolumn{1}{||r}{\textit{16}} & \textit{35} & \textit{53} & \textit{%
51} & \textit{39} & \textit{48} & \textit{43} & \textit{22} & \textit{13} &
\textit{6} & \multicolumn{1}{||r|}{326} \\
4 & \multicolumn{1}{||r}{\textit{3}} & \textit{29} & \textit{62} & \textit{61%
} & \textit{40} & \textit{41} & \textit{43} & \textit{32} & \textit{9} &
\textit{6} & \multicolumn{1}{||r|}{326} \\
5 & \multicolumn{1}{||r}{\textit{2}} & \textit{24} & \textit{91} & \textit{54%
} & \textit{39} & \textit{43} & \textit{23} & \textit{24} & \textit{16} &
\textit{10} & \multicolumn{1}{||r|}{326} \\
6 & \multicolumn{1}{||r}{0} & 0 & 0 & 0 & \textbf{70} & \textbf{65} &
\textbf{51} & \textbf{60} & \textbf{45} & \textbf{35} &
\multicolumn{1}{||r|}{326} \\
7 & \multicolumn{1}{||r}{0} & 0 & 0 & 0 & \textbf{53} & \textbf{36} &
\textbf{52} & \textbf{74} & \textbf{56} & \textbf{55} &
\multicolumn{1}{||r|}{326} \\
8 & \multicolumn{1}{||r}{0} & 0 & 0 & 0 & \textbf{35} & \textbf{33} &
\textbf{33} & \textbf{57} & \textbf{80} & \textbf{88} &
\multicolumn{1}{||r|}{326} \\
9 & \multicolumn{1}{||r}{0} & 0 & 0 & 0 & \textbf{10} & \textbf{23} &
\textbf{35} & \textbf{40} & \textbf{95} & \textbf{123} &
\multicolumn{1}{||r|}{326} \\ \cline{1-11}\cline{2-12}
$\beta $ & \multicolumn{1}{|c}{\textbf{0.65}} & \multicolumn{1}{|l|}{\textbf{%
1.60}} & \multicolumn{1}{|r}{\textbf{3.04}} & \multicolumn{1}{|r}{\textbf{%
2.71}} & \multicolumn{1}{|r}{5.25} & \multicolumn{1}{|l|}{5.25} &
\multicolumn{1}{|r}{5.39} & \multicolumn{1}{|r}{6.26} & \multicolumn{1}{|r}{
7.17} & \multicolumn{1}{|r|}{7.68} &  \\ \hline
\end{tabular}}

\bigskip
{\tiny\begin{tabular}{|l|lrrrrrrrrr|r|}
\multicolumn{12}{l}{\textbf{Table 9:\ M}$_{1,4},$ c\textbf{ontingency table
of first-order marginals of }$cohC_{1}(4).$} \\ \hline
\multicolumn{1}{|l|}{Borda} & \multicolumn{11}{|c|}{items} \\ \cline{2-12}
\multicolumn{1}{|l|}{scores} & j10 & j7 & j4 & j9 & j3 & j1 & j2 & j6 & j5 &
\multicolumn{1}{r|}{j8} & sum \\ \hline
0 & \multicolumn{1}{||r}{\textbf{164}} & \textbf{93} & \textbf{44} & \textbf{%
14} & 0 & 0 & 0 & 0 & 0 & 0 & \multicolumn{1}{||r|}{315} \\
1 & \multicolumn{1}{||r}{\textit{78}} & \textit{71} & \textit{30} & \textit{%
36} & \textit{10} & \textit{31} & \textit{32} & \textit{9} & \textit{16} &
\multicolumn{1}{r|}{\textit{2}} & \multicolumn{1}{||r|}{315} \\
2 & \multicolumn{1}{||r}{\textit{44}} & \textit{53} & \textit{49} & \textit{%
50} & \textit{32} & \textit{39} & \textit{27} & \textit{10} & \textit{8} &
\textit{3} & \multicolumn{1}{||r|}{315} \\
3 & \multicolumn{1}{||r}{\textit{22}} & \textit{52} & \textit{58} & \textit{%
87} & \textit{24} & \textit{20} & \textit{24} & \textit{15} & \textit{7} &
\multicolumn{1}{r|}{\textit{6}} & \multicolumn{1}{||r|}{315} \\
4 & \multicolumn{1}{||r}{\textit{5}} & \textit{17} & \textit{35} & \textit{43%
} & \textit{51} & \textit{61} & \textit{41} & \textit{25} & \textit{23} &
\textit{14} & \multicolumn{1}{||r|}{315} \\
5 & \multicolumn{1}{||r}{\textit{1}} & \textit{11} & \textit{61} & \textit{46%
} & \textit{43} & \textit{42} & \textit{34} & \textit{35} & \textit{26} &
\textit{16} & \multicolumn{1}{||r|}{315} \\
6 & \multicolumn{1}{||r}{\textit{1}} & \textit{18} & \textit{38} & \textit{39%
} & \textit{52} & \textit{37} & \textit{37} & \textit{49} & \textit{28} &
\textit{16} & \multicolumn{1}{||r|}{315} \\
7 & \multicolumn{1}{||r}{0} & 0 & 0 & 0 & \textbf{49} & \textbf{44} &
\textbf{51} & \textbf{61} & \textbf{54} & \textbf{56} &
\multicolumn{1}{||r|}{315} \\
8 & \multicolumn{1}{||r}{0} & 0 & 0 & 0 & \textbf{37} & \textbf{28} &
\textbf{47} & \textbf{72} & \textbf{69} & \textbf{52} &
\multicolumn{1}{||r|}{315} \\
9 & \multicolumn{1}{||r}{0} & 0 & 0 & 0 & \textbf{17} & \textbf{13} &
\textbf{22} & \textbf{39} & \textbf{84} & \textbf{140} &
\multicolumn{1}{||r|}{315} \\ \cline{1-11}\cline{2-12}
$\beta $ & \multicolumn{1}{|c}{\textbf{0.83}} & \multicolumn{1}{|l|}{\textbf{%
1.79}} & \multicolumn{1}{|r}{\textbf{3.10}} & \multicolumn{1}{|r}{\textbf{%
3.28}} & \multicolumn{1}{|r}{5.30} & \multicolumn{1}{|l|}{4.74} &
\multicolumn{1}{|r}{5.22} & \multicolumn{1}{|r}{6.34} & \multicolumn{1}{|r}{
6.76} & \multicolumn{1}{|r|}{7.64} &  \\ \hline
\end{tabular}}
\bigskip

{\tiny\begin{tabular}{|l|lrrrrrrrrr|r|}
\multicolumn{12}{l}{\textbf{Table 10:\ M}$_{1,5},$ c\textbf{ontingency table
of first-order marginals of }$cohC_{1}(5).$} \\ \hline
\multicolumn{1}{|l|}{Borda} & \multicolumn{11}{|c|}{items} \\ \cline{2-12}
\multicolumn{1}{|l|}{scores} & j10 & j7 & j4 & j9 & j3 & j1 & j2 & j6 & j5 &
j8 & sum \\ \hline
0 & \multicolumn{1}{||r}{\textit{188}} & \textit{99} & \textit{36} & \textit{%
10} & \textit{6} & \textit{25} & \textit{30} & \textit{22} & \textit{34} &
\textit{2} & \multicolumn{1}{||r|}{452} \\
1 & \multicolumn{1}{||r}{\textit{132}} & \textit{109} & \textit{69} &
\textit{57} & \textit{12} & \textit{30} & \textit{21} & \textit{13} &
\textit{9} & \textit{0} & \multicolumn{1}{||r|}{452} \\
2 & \multicolumn{1}{||r}{\textit{69}} & \textit{88} & \textit{59} & \textit{%
67} & \textit{28} & \textit{46} & \textit{40} & \textit{28} & \textit{20} &
\textit{7} & \multicolumn{1}{||r|}{452} \\
3 & \multicolumn{1}{||r}{\textit{39}} & \textit{72} & \textit{85} & \textit{%
92} & \textit{34} & \textit{44} & \textit{21} & \textit{31} & \textit{25} &
\textit{9} & \multicolumn{1}{||r|}{452} \\
4 & \multicolumn{1}{||r}{\textit{12}} & \textit{35} & \textit{76} & \textit{%
81} & \textit{50} & \textit{57} & \textit{53} & \textit{36} & \textit{38} &
\textit{14} & \multicolumn{1}{||r|}{452} \\
5 & \multicolumn{1}{||r}{\textit{6}} & \textit{29} & \textit{63} & \textit{72%
} & \textit{63} & \textit{64} & \textit{53} & \textit{41} & \textit{40} &
\textit{21} & \multicolumn{1}{||r|}{452} \\
6 & \multicolumn{1}{||r}{\textit{3}} & \textit{11} & \textit{34} & \textit{36%
} & \textit{71} & \textit{68} & \textit{64} & \textit{75} & \textit{45} &
\textit{45} & \multicolumn{1}{||r|}{452} \\
7 & \multicolumn{1}{||r}{\textit{3}} & \textit{9} & \textit{30} & \textit{37}
& \textit{87} & \textit{45} & \textit{62} & \textit{73} & \textit{57} &
\textit{49} & \multicolumn{1}{||r|}{452} \\
8 & \multicolumn{1}{||r}{0} & 0 & 0 & 0 & \textbf{71} & \textbf{41} &
\textbf{47} & \textbf{72} & \textbf{95} & \textbf{126} &
\multicolumn{1}{||r|}{452} \\
9 & \multicolumn{1}{||r}{0} & 0 & 0 & 0 & \textbf{30} & \textbf{32} &
\textbf{61} & \textbf{61} & \textbf{89} & \textbf{179} &
\multicolumn{1}{||r|}{452} \\ \cline{1-11}\cline{2-2}\cline{6-12}
$\beta $ & \multicolumn{1}{|c}{\textbf{1.12}} & \multicolumn{1}{|l|}{\textbf{%
2.02}} & \multicolumn{1}{|r}{\textbf{3.26}} & \multicolumn{1}{|r}{\textbf{%
3.60}} & \multicolumn{1}{|r}{5.70} & \multicolumn{1}{|l|}{4.74} &
\multicolumn{1}{|r}{5.27} & \multicolumn{1}{|r}{5.75} & \multicolumn{1}{|r}{
5.99} & \multicolumn{1}{|r|}{7.60} &  \\ \hline
\end{tabular}}
\bigskip

{\tiny\begin{tabular}{|l|lrrrrrrrrr|r|}
\multicolumn{12}{l}{\textbf{Table 11:\ M}$_{1,6},$ c\textbf{ontingency table
of first-order marginals of }$cohC_{1}(6).$} \\ \hline
\multicolumn{1}{|l|}{Borda} & \multicolumn{11}{|c|}{items} \\ \cline{2-12}
\multicolumn{1}{|l|}{scores} & j10 & j7 & j4 & j9 & j3 & j1 & j2 & j6 & j5 &
\multicolumn{1}{r|}{j8} & sum \\ \hline
0 & \multicolumn{1}{||r}{\textit{151}} & \textit{81} & \textit{31} & \textit{%
14} & \textit{8} & \textit{14} & \textit{19} & \textit{18} & \textit{39} &
\textit{0} & \multicolumn{1}{||r|}{375} \\
1 & \multicolumn{1}{||r}{\textit{112}} & \textit{79} & \textit{44} & \textit{%
33} & \textit{12} & \textit{21} & \textit{26} & \textit{25} & \textit{19} &
\textit{4} & \multicolumn{1}{||r|}{375} \\
2 & \multicolumn{1}{||r}{\textit{66}} & \textit{72} & \textit{52} & \textit{%
63} & \textit{16} & \textit{24} & \textit{29} & \textit{22} & \textit{28} &
\textit{3} & \multicolumn{1}{||r|}{375} \\
3 & \multicolumn{1}{||r}{\textit{26}} & \textit{52} & \textit{68} & \textit{%
68} & \textit{22} & \textit{45} & \textit{31} & \textit{29} & \textit{25} &
\textit{9} & \multicolumn{1}{||r|}{375} \\
4 & \multicolumn{1}{||r}{\textit{8}} & \textit{26} & \textit{42} & \textit{37%
} & \textit{52} & \textit{67} & \textit{41} & \textit{45} & \textit{45} &
\textit{12} & \multicolumn{1}{||r|}{375} \\
5 & \multicolumn{1}{||r}{\textit{8}} & \textit{27} & \textit{56} & \textit{61%
} & \textit{44} & \textit{49} & \textit{42} & \textit{36} & \textit{28} &
\textit{24} & \multicolumn{1}{||r|}{375} \\
6 & \multicolumn{1}{||r}{\textit{3}} & \textit{21} & \textit{36} & \textit{52%
} & \textit{64} & \textit{42} & \textit{49} & \textit{50} & \textit{29} &
\textit{29} & \multicolumn{1}{||r|}{375} \\
7 & \multicolumn{1}{||r}{\textit{0}} & \textit{7} & \textit{25} & \textit{31}
& \textit{70} & \textit{43} & \textit{44} & \textit{59} & \textit{45} &
\textit{51} & \multicolumn{1}{||r|}{375} \\
8 & \multicolumn{1}{||r}{\textit{1}} & \textit{10} & \textit{21} & \textit{16%
} & \textit{66} & \textit{33} & \textit{46} & \textit{49} & \textit{45} &
\textit{88} & \multicolumn{1}{||r|}{375} \\
9 & \multicolumn{1}{||r}{0} & 0 & 0 & 0 & \textbf{21} & \textbf{37} &
\textbf{48} & \textbf{42} & \textbf{72} & \textbf{155} &
\multicolumn{1}{||r|}{375} \\ \cline{1-11}\cline{2-2}\cline{6-12}
$\beta $ & \multicolumn{1}{|c}{\textbf{1.12}} & \multicolumn{1}{|l|}{\textbf{%
2.33}} & \multicolumn{1}{|r}{\textbf{3.62}} & \multicolumn{1}{|r}{\textbf{%
3.93}} & \multicolumn{1}{|r}{5.68} & \multicolumn{1}{|l|}{4.98} &
\multicolumn{1}{|r}{5.21} & \multicolumn{1}{|r}{5.33} & \multicolumn{1}{|r}{
5.25} & \multicolumn{1}{|r|}{7.56} &  \\ \hline
\end{tabular}}
\bigskip

{\tiny\begin{tabular}{|l|lrrrrrrrrr|r|}
\multicolumn{12}{l}{\textbf{Table 12:\ M}$_{1,7},$ c\textbf{ontingency table
of first-order marginals of }$cohC_{1}(7).$} \\ \hline
\multicolumn{1}{|l|}{Borda} & \multicolumn{11}{|c|}{items} \\ \cline{2-12}
\multicolumn{1}{|l|}{scores} & j10 & j7 & j4 & j9 & j3 & j1 & j2 & j6 & j5 &
j8 & sum \\ \hline
0 & \multicolumn{1}{||r}{\textit{129}} & \textit{65} & \textit{46} & \textit{%
14} & \textit{11} & \textit{24} & \textit{35} & \textit{23} & \textit{52} &
\textit{2} & \multicolumn{1}{||r|}{401} \\
1 & \multicolumn{1}{||r}{\textit{122}} & \textit{77} & \textit{53} & \textit{%
35} & \textit{14} & \textit{28} & \textit{24} & \textit{19} & \textit{25} &
\textit{4} & \multicolumn{1}{||r|}{401} \\
2 & \multicolumn{1}{||r}{\textit{74}} & \textit{69} & \textit{50} & \textit{%
51} & \textit{24} & \textit{41} & \textit{36} & \textit{31} & \textit{19} &
\textit{6} & \multicolumn{1}{||r|}{401} \\
3 & \multicolumn{1}{||r}{\textit{36}} & \textit{51} & \textit{31} & \textit{%
66} & \textit{44} & \textit{48} & \textit{39} & \textit{46} & \textit{30} &
\textit{10} & \multicolumn{1}{||r|}{401} \\
4 & \multicolumn{1}{||r}{\textit{24}} & \textit{50} & \textit{40} & \textit{%
71} & \textit{51} & \textit{45} & \textit{38} & \textit{37} & \textit{27} &
\textit{18} & \multicolumn{1}{||r|}{401} \\
5 & \multicolumn{1}{||r}{\textit{7}} & \textit{45} & \textit{49} & \textit{56%
} & \textit{43} & \textit{53} & \textit{39} & \textit{48} & \textit{32} &
\textit{29} & \multicolumn{1}{||r|}{401} \\
6 & \multicolumn{1}{||r}{\textit{5}} & \textit{23} & \textit{73} & \textit{68%
} & \textit{42} & \textit{50} & \textit{42} & \textit{33} & \textit{40} &
\textit{25} & \multicolumn{1}{||r|}{401} \\
7 & \multicolumn{1}{||r}{\textit{3}} & \textit{10} & \textit{31} & \textit{28%
} & \textit{85} & \textit{39} & \textit{43} & \textit{54} & \textit{51} &
\textit{57} & \multicolumn{1}{||r|}{401} \\
8 & \multicolumn{1}{||r}{\textit{1}} & \textit{3} & \textit{17} & \textit{5}
& \textit{58} & \textit{46} & \textit{47} & \textit{65} & \textit{57} &
\textit{102} & \multicolumn{1}{||r|}{401} \\
9 & \multicolumn{1}{||r}{\textit{0}} & \textit{8} & \textit{11} & \textit{7}
& \textit{29} & \textit{27} & \textit{58} & \textit{45} & \textit{68} &
\textit{148} & \multicolumn{1}{||r|}{401} \\
\cline{1-11}\cline{2-2}\cline{6-12}
$\beta $ & \multicolumn{1}{|c}{\textbf{1.42}} & \multicolumn{1}{|l|}{\textbf{%
2.74}} & \multicolumn{1}{|r}{\textbf{3.84}} & \multicolumn{1}{|r}{\textbf{%
4.00}} & \multicolumn{1}{|r}{5.45} & \multicolumn{1}{|l|}{4.70} &
\multicolumn{1}{|r}{5.02} & \multicolumn{1}{|r}{5.26} & \multicolumn{1}{|r}{
5.20} & \multicolumn{1}{|r|}{7.38} &  \\ \hline
\end{tabular}}

\textbf{Example 6:} Tables 6 to 12 display \textbf{M}$_{1,\alpha }$ for $%
\alpha =1,...,7,$ the contingency tables of first order marginals of the
seven coherent clusters of the SUSHI data, respectively. We observe the
following:

Each one of them reveals the nature of the riffle shuffles of its coherent
cluster, which are summarized in Table 13. The number of observed $(4,6)$
blocks of scores for the seven coherent clusters, ($\tau
_{J_{1}}(S_{1}),\tau _{J_{2}}(S_{2})),$ is only 27 in Table 13 out of the
possible total number of $10!/(4!6!)=210$. The counts of the observed $(4,6)$
blocks do not seem to be uniformly distributed in Table 13. Furthermore, we
observe that as $\alpha $ increases from 1 to 7, the magnitude of riffle
shuffles, $T_{v}(\tau _{J_{1}}(S_{1})),$ increases in the coherent clusters
from 6 to 12. Integers in bold in Table 13 are the shuffled-crossed
scores.\bigskip

{\tiny\begin{tabular}{|llrr||l|lrr|}
\multicolumn{8}{l}{\textbf{Table 13: Types of riffle shuffles in the 7
coherent clusters of SUSHI\ data.}} \\ \hline
$cohC_{1}(\alpha )$ & \multicolumn{1}{|l}{scores given to} & sum of &  & $%
cohC_{1}(\alpha )$ & scores given to & sum of &  \\
& \multicolumn{1}{|l}{$\left\{ j10,j7,j4,j9\right\} $} & scores & count &  &
$\left\{ j10,j7,j4,j9\right\} $ & scores & count \\ \hline\hline
$cohC_{1}(1)$ & \multicolumn{1}{|c}{$\left\{ 0,1,2,3\right\} $} &
\multicolumn{1}{c}{6} & 314 & \multicolumn{1}{||r|}{$cohC_{1}(6)$} &
\multicolumn{1}{|c}{$\left\{ 0,1,2,\mathbf{8}\right\} $} &
\multicolumn{1}{c}{11} & 48 \\ \cline{1-4}
$cohC_{1}(2)$ & \multicolumn{1}{|c}{$\left\{ 0,1,2,\mathbf{4}\right\} $} &
\multicolumn{1}{c}{7} & 235 & \multicolumn{1}{||r|}{} & \multicolumn{1}{|c}{$%
\left\{ 0,1,\mathbf{7},3\right\} $} & \multicolumn{1}{c}{11} & 63 \\
\cline{1-4}
$cohC_{1}(3)$ & \multicolumn{1}{|c}{$\left\{ 0,1,2,\mathbf{5}\right\} $} &
\multicolumn{1}{c}{8} & 171 & \multicolumn{1}{||r|}{} & \multicolumn{1}{|c}{$%
\left\{ 0,\mathbf{6},2,3\right\} $} & \multicolumn{1}{c}{11} & 53 \\
& \multicolumn{1}{|c}{$\left\{ 0,1,\mathbf{4},3\right\} $} &
\multicolumn{1}{c}{8} & 155 & \multicolumn{1}{||r|}{} & \multicolumn{1}{|c}{$%
\left\{ \mathbf{5},1,2,3\right\} $} & \multicolumn{1}{c}{11} & 98 \\
\cline{1-4}
$cohC_{1}(4)$ & \multicolumn{1}{|c}{$\left\{ 0,1,2,\mathbf{6}\right\} $} &
\multicolumn{1}{c}{9} & 96 & \multicolumn{1}{||r|}{} & \multicolumn{1}{|c}{$%
\left\{ 0,1,\mathbf{4,6}\right\} $} & \multicolumn{1}{c}{11} & 59 \\
& \multicolumn{1}{|c}{$\left\{ 0,1,\mathbf{5},3\right\} $} &
\multicolumn{1}{c}{9} & 119 & \multicolumn{1}{||r|}{} & \multicolumn{1}{|c}{$%
\left\{ 0,\mathbf{4},2,\mathbf{5}\right\} $} & \multicolumn{1}{c}{11} & 54
\\ \cline{5-8}
& \multicolumn{1}{|c}{$\left\{ 0,\mathbf{4},2,3\right\} $} &
\multicolumn{1}{c}{9} & 100 & \multicolumn{1}{||r|}{$cohC_{1}(7)$} &
\multicolumn{1}{|c}{$\left\{ 0,1,2,\mathbf{9}\right\} $} &
\multicolumn{1}{c}{12} & 26 \\ \cline{1-4}
$cohC_{1}(5)$ & \multicolumn{1}{|c}{$\left\{ 0,1,2,\mathbf{7}\right\} $} &
\multicolumn{1}{c}{10} & 79 & \multicolumn{1}{||r|}{} & \multicolumn{1}{|c}{$%
\left\{ 0,1,\mathbf{8},3\right\} $} & \multicolumn{1}{c}{12} & 26 \\
& \multicolumn{1}{|c}{$\left\{ 0,1,\mathbf{6},3\right\} $} &
\multicolumn{1}{c}{10} & 84 & \multicolumn{1}{||r|}{} & \multicolumn{1}{|c}{$%
\left\{ 0,\mathbf{7},2,3\right\} $} & \multicolumn{1}{c}{12} & 33 \\
& \multicolumn{1}{|c}{$\left\{ 0,\mathbf{5},2,3\right\} $} &
\multicolumn{1}{c}{10} & 85 & \multicolumn{1}{||r|}{} & \multicolumn{1}{|c}{$%
\left\{ \mathbf{6},1,2,3\right\} $} & \multicolumn{1}{c}{12} & 43 \\
& \multicolumn{1}{|c}{$\left\{ \mathbf{4},1,2,3\right\} $} &
\multicolumn{1}{c}{10} & 119 & \multicolumn{1}{||r|}{} & \multicolumn{1}{|c}{%
$\left\{ 0,\mathbf{4,5},3\right\} $} & \multicolumn{1}{c}{12} & 38 \\
& \multicolumn{1}{|c}{$\left\{ 0,1,\mathbf{4,5}\right\} $} &
\multicolumn{1}{c}{10} & 85 & \multicolumn{1}{||r|}{} & \multicolumn{1}{|c}{$%
\left\{ 0,\mathbf{4},2,\mathbf{6}\right\} $} & \multicolumn{1}{c}{12} & 39
\\ \cline{1-4}
&  &  &  & \multicolumn{1}{||r|}{} & \multicolumn{1}{|c}{$\left\{ 0,1,%
\mathbf{4},\mathbf{7}\right\} $} & \multicolumn{1}{c}{12} & 49 \\
& \multicolumn{1}{c}{} & \multicolumn{1}{c}{} &  & \multicolumn{1}{||r|}{} &
\multicolumn{1}{|c}{$\left\{ 0,1,\mathbf{5,6}\right\} $} &
\multicolumn{1}{c}{12} & 82 \\
&  &  &  & \multicolumn{1}{||r|}{} & \multicolumn{1}{|c}{$\left\{ \mathbf{4}%
,1,2,\mathbf{5}\right\} $} & \multicolumn{1}{c}{12} & 65 \\ \hline
\end{tabular}}
\bigskip

The counts in Table 13 are calculated from \textbf{M}$_{1,\alpha }$ for $%
\alpha =1,...,7,$ by reasoning on the permutation of scores between the sets
$S_{1}$ and $S_{2}$. Here are the details, where $J_{1}=\left\{
j10,j7,j4,j9\right\} $.

a)$\ cohC_{1}(1)$

$|\left\{ 0,1,2,3\right\} |=314,$ which is the number of $0$s attributed to $%
J_{1}$ in \textbf{M}$_{1,1}.\ $Among the \textbf{M}$_{1,\alpha }$ for $%
\alpha =1,...,7$, note that \textbf{M}$_{1,1}$ is the only contingency table
of first-order marginals which is block diagonal.

b) $cohC_{1}(2)$

$|\left\{ 0,1,2,4\right\} |=235,$ which is the number of $4$s attributed to $%
J_{1}$ in \textbf{M}$_{1,2}.$

c) $cohC_{1}(3)$

$|\left\{ 0,1,2,5\right\} |=171,$ which is the number of $5$s attributed to $%
J_{1}\ $in \textbf{M}$_{1,3}.$

$|\left\{ 0,1,\mathbf{4},3\right\} |=155,$ which is the number of $4$s
attributed to $J_{1}$ in \textbf{M}$_{1,3}.$

d) $cohC_{1}(4)$

$|\left\{ 0,1,2,\mathbf{6}\right\} |=96,$ which is the number of $6$s
attributed to $J_{1}$ in \textbf{M}$_{1,4}.$

$|\left\{ 0,1,\mathbf{5},3\right\} |=119,$ which is the number of $5$s
attributed to $J_{1}$ in \textbf{M}$_{1,4}.$

$|\left\{ 0,\mathbf{4},2,3\right\} |=100,$ which is the number of $4$s
attributed to $J_{1}$ in \textbf{M}$_{1,4}.$

e) $cohC_{1}(5)$

$|\left\{ 0,1,2,\mathbf{7}\right\} |=79,$ which is the number of $7$s
attributed to $J_{1}$ in \textbf{M}$_{1,5}.$

$|\left\{ 0,1,\mathbf{6},3\right\} |=84,$ which is the number of $6$s
attributed to $J_{1}$ in \textbf{M}$_{1,5}.$

$|\left\{ 0,\mathbf{5},2,3\right\} |=85,$ which is the number of $1$s not
attributed to $J_{1}$ in \textbf{M}$_{1,5}.$

$|\left\{ 0,1,\mathbf{4},\mathbf{5}\right\} |+|\left\{ 0,\mathbf{5}%
,2,3\right\} |=170,$ which is the total number of $5$s attributed to $J_{1}\
$in $\mathbf{M}_{1,5};$ so $|\left\{ 0,1,\mathbf{4},\mathbf{5}\right\}
|=170-85=85.$

$|\left\{ \mathbf{4},1,2,3\right\} |=119,$ which is the number of $0$s not
attributed to $J_{1}\ $in $\mathbf{M}_{1,5}.$

f) $cohC_{1}(6)$

$|\left\{ 0,1,2,\mathbf{8}\right\} |=48,$ which is the number of $8$s
attributed to $J_{1}\ $in $\mathbf{M}_{1,6}.$

$|\left\{ 0,1,\mathbf{7},3\right\} |=63,$ which is the number of $7$s
attributed to $J_{1}\ $in $\mathbf{M}_{1,6}.$

$|\left\{ \mathbf{5},1,2,3\right\} |=98,$ which is the number of $0$s not
attributed to $J_{1}\ $in $\mathbf{M}_{1,6}.$

$|\left\{ 0,\mathbf{4},2,\mathbf{5}\right\} |=152-98=54,$ where $152$ is the
total number of $5$s attributed to $J_{1}\ $in $\mathbf{M}_{1,6}.$

$|\left\{ 0,1,\mathbf{4},\mathbf{6}\right\} |=113-54=59,$ where $113$ is the
total number of $4$s attributed to $J_{1}\ $in $\mathbf{M}_{1,6}.$

$|\left\{ 0,\mathbf{6},2,3\right\} |=112-59=53,$ where $112$ is the total
number of $6$s attributed to $J_{1}\ $in $\mathbf{M}_{1,6}.$

g) $cohC_{1}(7)$

$|\left\{ 0,1,2,\mathbf{9}\right\} |=26,$ which is the number of $9$s
attributed to $J_{1}\ $in $\mathbf{M}_{1,7}.$

$|\left\{ 0,1,\mathbf{8},3\right\} |=26,$ which is the number of $8$s
attributed to $J_{1}\ $in $\mathbf{M}_{1,7}.$

For the remaining counts, we have to solve the following system of 7 linear
equations, where, $u=|\left\{ 0,\mathbf{7},2,3\right\} |$, $t=|\left\{ 0,%
\mathbf{4},\mathbf{5},3\right\} |$, $s=|\left\{ 0,\mathbf{4},2,\mathbf{6}%
\right\} |$, $w=|\left\{ 0,1,\mathbf{4,7}\right\} |$, $z=|\left\{ 0,1,%
\mathbf{5,6}\right\} |$, $x=|\left\{ \mathbf{6},1,2,3\right\} |$, and $%
y=|\left\{ \mathbf{4},1,2,5\right\} |$.

$x+y=147,\ $which is the number of $0$s not attributed to $J_{1}\ $in $%
\mathbf{M}_{1,7}.$

$u+w=72,\ $which is the number of $7$s attributed to $J_{1}\ $in $\mathbf{M}%
_{1,7}.$

$s+z+x=169,\ $which is the number of $6$s attributed to $J_{1}\ $in $\mathbf{%
M}_{1,7}.$

$t+z+y=157,\ $which is the number of $5$s attributed to $J_{1}\ $in $\mathbf{%
M}_{1,7}.$

$t+s+w+y=185,\ $which is the number of $4$s attributed to $J_{1}\ $in $%
\mathbf{M}_{1,7}.$

$u+t+x=158,\ $which is the number of $3$s attributed to $J_{1}\ $in $\mathbf{%
M}_{1,7}.$

$u+s+x+y=218,\ $which is the number of $2$s attributed to $J_{1}\ $in $%
\mathbf{M}_{1,7}.$

\section{Crossing index}

The following $(d_{1},d_{2})$ crossing index is based on the internal
dispersion of a voting profile. \bigskip

\textbf{Definition 3}: For a voting profile $V$ we define its crossing index
to be

\begin{eqnarray*}
Cross(V) &=&1-\frac{\delta _{1}(V_{d_{1},d_{2}})}{\max_{V}\delta
_{1}(V_{d_{1},d_{2}})}, \\
&=&1-\frac{\delta _{1}(V_{d_{1},d_{2}})}{2\frac{d_{1}d_{2}}{d(d-1)}}\ \
\text{by Proposition 2.}
\end{eqnarray*}%
where $\delta _{1}(V_{d_{1},d_{2}})$ is the first taxicab dispersion
obtained from TCA of $V$ and $(d_{1},d_{2})$ represents the optimal TCA
binary partition of the $d$ items of $V$ such that $d=d_{1}+d_{2}.\bigskip $

\textbf{Proposition 4}: The crossing index of a coherent cluster is%
\begin{equation*}
Cross(cohC(\alpha ))=\frac{2(\alpha -1)}{d_{1}d_{2}}.
\end{equation*}%
\bigskip

\textbf{Example 7:} The last column in Table 3 contains the values of the
crossing indices of the seven coherent clusters of the first iteration of
SUSHI data. We observe: a) $Cross(cohC_{1}(1))=0$, because the structure of
its matrix of first order marginals, M$_{1,1},$ is block diagonal; which
means that the permutation $\tau $ is the identical permutation, so there
are no crossing of scores between the two subsets of items $J_{1}$ and $J_{2}
$ in $cohC_{1}(1).$\ b) $Cross(cohC_{1}(\alpha ))$ for $\alpha =1,...,7$ is
a uniformly increasing function of $\alpha ,$ similar in spirit to the $%
T_{v}(\tau _{J_{1}}(S_{1}))$ statistic. c) For the incoherent cluster $%
V_{1,8}$, we have: $\delta _{1}(V_{1,8})=0.2354$ given in Example 3; and $%
d_{1}=d_{2}=5$ from Figure 10. So $Cross(V_{1,8})=1-\frac{0.2354}{%
2(5)(5)/(10(9))}=1-0.4237=0.5763.$

\section{Coherent group}

Our aim is to explore a given voting profile $V$ by uncovering its coherent
mixture groups, see equation (1); that is, $V=\cup _{g=1}^{G}cohG(g)\cup
noisyG$, where $G$ represents the number of coherent groups and $cohG(g)$ is
the $g$th coherent group. The computation is done by an iterative procedure
in $n_{G}$ steps for $n_{G}\geq G$ that we describe: \bigskip

For $g=1$; let $V_{1}=V;$ compute $cohG(1)$ from $V_{1},$ then partition $%
V_{1}=V_{2}\cup cohG(1);$

For $g=2$; compute $cohG(2)$ from $V_{2},$ then partition $V_{2}=V_{3}\cup
cohG(2);$

By continuing the above procedure, after $n_{G}$ steps, we get $V=\cup
_{g=1}^{n_{G}}cohG(g).\bigskip $

However, some of the higher ordered coherent groups may have relatively
small sample sizes; so by considering these as outliers, we lump them
together thus forming the noisy group denoted by $noisyG$ in equation (1).

Let us recall the definition of a coherent group given in equation 2%
\begin{equation*}
cohG(g)=\cup _{\alpha =1}^{c_{g}}cohC_{g}(\alpha )\text{ \ for }g=1,...,G;
\end{equation*}%
that is, a coherent group is the union of its coherent clusters. This
implies that the sample size of $cohG(g)$ equals the sum of the sample sizes
of its coherent clusters
\begin{equation*}
|cohG(g)|\ =\sum_{\alpha =1}^{c_{g}}|cohC_{g}(\alpha )|.
\end{equation*}

As an example, for the SUSHI data, from the 2nd column of Table 3 we can
compute the sample size of the first coherent group
\begin{eqnarray*}
|cohG(1)| &=&\sum_{\alpha =1}^{c_{g}=7}|cohC_{1}(\alpha )| \\
&=&2418.
\end{eqnarray*}%
Furthermore, $cohG(1)$ is composed of 27 observed riffle shuffles summarized
in Table 13, which provides quite a detailed view of its inner structure.

The next result shows important characteristics of a coherent group
inherited from its coherent clusters.\bigskip

\textbf{Theorem 2}: ( Properties of a coherent group $cohG(g)$)

a) The first principal column factor score $\mathbf{g}_{1}$ of the $d$ items
in a coherent group is the weighted average of the first principal column
factor score $\mathbf{g}_{1}$ of the $d$ items of its coherent clusters;
that is,
\begin{eqnarray*}
g_{1}(j &\in &cohG(g))=\sum_{\alpha =1}^{c_{g}}\frac{|cohC_{g}(\alpha )|}{%
|cohG(g)|}g_{1}(j\in cohC_{g}(\alpha ))\text{\ \ for }j=1,...,d. \\
&=&\frac{2}{d-1}\sum_{\alpha =1}^{c_{g}}\frac{|cohC_{g}(\alpha )|}{|cohG(g)|}%
\beta (j\in cohC_{g}(\alpha ))-1\text{ \ \ by Proposition 3.}
\end{eqnarray*}%
And $corr(\mathbf{g}_{1}(cohG(g),\mathbf{\beta }(cohG(g))=1.$

b)\ The first TCA dispersion value of a coherent group is the weighted
average of the first TCA dispersion values of its coherent clusters; that
is,
\begin{equation*}
\delta _{1}(cohG(g))=\sum_{\alpha =1}^{c_{g}}\frac{|cohC_{g}(\alpha )|}{%
|cohG(g)|}\delta _{1}(cohC_{g}(\alpha )).
\end{equation*}

c) The crossing index of a coherent group is the weighted average of the
crossing indices of its coherent clusters; that is,
\begin{equation*}
Cross(cohG(g))=\sum_{\alpha =1}^{c_{g}}\frac{|cohC_{g}(\alpha )|}{|cohG(g)|}%
Cross(cohC_{g}(\alpha )).
\end{equation*}%
\bigskip

\textbf{Example 8:} Table 14 summarizes the first four coherent groups of
SUSHI data, which emerged after 5 iterations. For $g=1$, we get $%
cohG(1)=\cup _{\alpha =1}^{c_{1}=7}cohC_{1}(\alpha );$ that is, the first
coherent group of voters, the majority, is composed of 48.36\% of the sample
with crossing index of 27.3\%. Standard errors of the Borda scale of the
items in $cohG(1)\ $in Table 14 are:
\begin{equation*}
(0.046,0.051,0.042,0.042,0.053,0.047,0.037,0.034,0.037,0.025).
\end{equation*}

We can discern the following grouped seriation (bucket ranking) of the items
\begin{equation*}
j8\succ j5\succ j6\succ \left\{ j3,j2\right\} \succ j1\succ \left\{
j9,j4\right\} \succ \left\{ j7\right\} \succ \left\{ j10\right\} .
\end{equation*}%
The groupings are based on the standard 95\% confidence intervals of the
Borda scale of the items.

The 2nd coherent group $cohG(2),$ summarized by its Borda scales in Table
14, is made up of eight coherent clusters; it is composed of 19.0\% of the
sample with crossing index of 35.38\%. The voters in this coherent group
disapprove\\ $\left\{ uni(seaurchin),sake(salmonroe)\right\} ,$ which are
considered more \textquotedblright daring sushis\textquotedblright .

The third coherent group $cohG(3),$ summarized by its Borda scales in Table
14, is made up of eight coherent clusters; it is composed of 13.24\% of the
sample with crossing index of 27.3\%. The voters in this coherent group
prefer the three types of tuna sushis with sea urchin sushis.

The fourth coherent group $cohG(4),$ summarized by its Borda scales in Table
14, is made up of eight coherent clusters; it is composed of 6.94\% of the
sample with crossing index of 35.27\%. The voters disapprove the three types
of tuna sushis.\bigskip

\textbf{Remark 6:}

a) Note that the number of preferred sushis in $cohG(1)$ and $cohG(2)$ are
six; that is $|J_{2}|=6.$ While the number of preferred sushis in $cohG(3)$
and $cohG(4)$ are four.

b) The four coherent groups summarized in Table 14 can also be described as
two bipolar latent factors: By noting that the only major difference between
the first two coherent groups is that (\textit{5. uni (sea urchin),\ 6. sake
(salmon roe))} are swapped with (\textit{7. tamago (egg), 4. ika (squid)).\ }%
While the only major difference between the third and fourth coherent groups
is that the three tunas are swapped with (\textit{4. ika (squid),\ 5. uni
(sea urchin), 1. ebi (shrimp)).}

c) We consider the fifth group as noisy (outliers not shown) composed of
12.36\% of the remaining sample: it contains $cohG(5)=\cup _{\alpha
=1}^{2}cohC_{5}(\alpha )$ whose sample size is $38$, a very small number.
For the sake of completeness we also provide the sample sizes of its two
coherent clusters $|cohC_{5}(1)|=22$ and $|cohC_{5}(2)|=16$.

\bigskip \bigskip
{\tiny \begin{tabular}{|lrlr|}
\multicolumn{4}{l|}{\textbf{Table 14: The first four coherent groups of
SUSHI data and related statistics.}} \\ \hline
$\mathbf{cohG(1)=\cup }_{\alpha =1}^{7}\mathbf{cohC}_{1}\mathbf{(\alpha )}$
& \multicolumn{1}{|r}{$\mathbf{\beta }$} & \multicolumn{1}{||l}{$\mathbf{%
cohG(2)=\cup }_{\alpha =1}^{8}\mathbf{cohC}_{2}\mathbf{(\alpha )}$} &
\multicolumn{1}{|r|}{$\mathbf{\beta }$} \\ \hline\hline
\textit{8. toro (fatty tuna)} & \multicolumn{1}{|r}{$\mathbf{7.62}$} &
\multicolumn{1}{||r|}{\textit{8. toro (fatty tuna)}} & \multicolumn{1}{|r|}{$%
\mathbf{6.15}$} \\
\textit{5. uni (sea urchin)} & \multicolumn{1}{|r}{$\mathbf{6.31}$} &
\multicolumn{1}{||r|}{\textit{2. anago (sea eel)}} & \multicolumn{1}{|r|}{$%
\mathbf{5.97}$} \\
\textit{6. sake (salmon roe)} & \multicolumn{1}{|r}{$\mathbf{5.92}$} &
\multicolumn{1}{||r|}{\textit{1. ebi (shrimp)}} & \multicolumn{1}{|r|}{$%
\mathbf{5.92}$} \\
\textit{3. maguro (tuna)} & \multicolumn{1}{|r|}{$\mathbf{5.49}$} &
\multicolumn{1}{||r|}{\textit{7. tamago (egg)}} & \multicolumn{1}{|r|}{$%
\mathbf{5.76}$} \\
\textit{2. anago (sea eel)} & \multicolumn{1}{|c||}{$\mathbf{5.35}$} &
\multicolumn{1}{||r|}{\textit{3. maguro (tuna)}} & \multicolumn{1}{|r|}{$%
\mathbf{5.55}$} \\
\textit{1. ebi (shrimp)} & \multicolumn{1}{|c||}{$\mathbf{5.04}$} &
\multicolumn{1}{||r|}{\textit{4. ika (squid)}} & \multicolumn{1}{|r|}{$%
\mathbf{5.41}$} \\ \hline
9. tekka-maki (tuna roll) & \multicolumn{1}{|r}{3.27} &
\multicolumn{1}{||r|}{9. tekka-maki (tuna roll)} & \multicolumn{1}{|r|}{3.80}
\\
4. ika (squid) & \multicolumn{1}{|r}{3.10} & \multicolumn{1}{||r|}{10.
kappa-maki (cucumber roll)} & \multicolumn{1}{|r|}{2.56} \\
7. tamago (egg) & \multicolumn{1}{|r}{1.94} & \multicolumn{1}{||r|}{6. sake
(salmon roe)} & \multicolumn{1}{|r|}{2.45} \\
10. kappa-maki (cucumber roll) & \multicolumn{1}{|r}{0.97} &
\multicolumn{1}{||r|}{5. uni (sea urchin)} & \multicolumn{1}{|r|}{1.44} \\
\hline\hline
$Cross(cohG(1))=27.3\%$ & \multicolumn{1}{|r}{} & \multicolumn{1}{||l}{$%
Cross(cohG(2))=35.38\%$} & \multicolumn{1}{|r|}{} \\
$|cohG(1)|=2418\ (48.36\%)$ & \multicolumn{1}{|r}{} & \multicolumn{1}{||l}{$%
|cohG(2)|=955\ (19.10\%)$} & \multicolumn{1}{|r|}{} \\ \hline\hline
&  &  &  \\ \hline\hline
$\mathbf{cohG(3)=\cup }_{\alpha =1}^{8}\mathbf{cohC}_{3}\mathbf{(\alpha )}$
& \multicolumn{1}{|r}{$\mathbf{\beta }$} & \multicolumn{1}{||l}{$\mathbf{%
cohG(4)=\cup }_{\alpha =1}^{8}\mathbf{cohC}_{4}\mathbf{(\alpha )}$} &
\multicolumn{1}{|r|}{$\mathbf{\beta }$} \\ \hline\hline
\textit{8. toro (fatty tuna)} & \multicolumn{1}{|r}{$\mathbf{7.31}$} &
\multicolumn{1}{||r}{\textit{4. ika (squid)}} & \multicolumn{1}{|r|}{$%
\mathbf{6.67}$} \\
\textit{6. sake (salmon roe)} & \multicolumn{1}{|r}{$\mathbf{6.62}$} &
\multicolumn{1}{||r}{\textit{5. uni (sea urchin)}} & \multicolumn{1}{|r|}{$%
\mathbf{6.50}$} \\
\textit{3. maguro (tuna)} & \multicolumn{1}{|r}{$\mathbf{6.30}$} &
\multicolumn{1}{||r}{\textit{6. sake (salmon roe)}} & \multicolumn{1}{|r|}{$%
\mathbf{6.43}$} \\
\textit{9. tekka-maki (tuna roll)} & \multicolumn{1}{|r}{$\mathbf{6.00}$} &
\multicolumn{1}{||r|}{\textit{1. ebi (shrimp)}} & \multicolumn{1}{|r|}{$%
\mathbf{6.16}$} \\ \hline
7. tamago (egg) & \multicolumn{1}{|c}{3.76} & \multicolumn{1}{||r}{8. toro
(fatty tuna)} & \multicolumn{1}{|r|}{3.69} \\
4. ika (squid) & \multicolumn{1}{|c}{3.41} & \multicolumn{1}{||r}{7. tamago
(egg)} & \multicolumn{1}{|r|}{3.39} \\
2. anago (sea eel) & \multicolumn{1}{|r}{3.00} & \multicolumn{1}{||r}{2.
anago (sea eel)} & \multicolumn{1}{|r|}{3.21} \\
1. ebi (shrimp) & \multicolumn{1}{|r}{2.92} & \multicolumn{1}{||r}{9.
tekka-maki (tuna roll)} & \multicolumn{1}{|r|}{3.14} \\
10. kappa-maki (cucumber roll) & \multicolumn{1}{|r}{2.86} &
\multicolumn{1}{||r}{10. kappa-maki (cucumber roll)} & \multicolumn{1}{|r|}{
2.99} \\
5. uni (sea urchin) & \multicolumn{1}{|r}{2.80} & \multicolumn{1}{||r}{3.
maguro (tuna)} & \multicolumn{1}{|r|}{2.80} \\ \hline\hline
$Cross(cohG(3))=31.37\%$ & \multicolumn{1}{|r}{} & \multicolumn{1}{||l}{$%
Cross(cohG(4))=35.27\%$} & \multicolumn{1}{|r|}{} \\ \hline
$|cohG(3)|\ =662\ (13.24\%)$ & \multicolumn{1}{|r}{} & \multicolumn{1}{||l}{$%
|cohG(4)|\ =347\ (6.94\%)$} & \multicolumn{1}{|r|}{} \\ \hline
\end{tabular}}

\section{APA data set}

The 1980 American Psychological Association (APA) presidential election had
five candidates: $\left\{ A,C\right\} $ were \textit{research}
psychologists, $\left\{ D,E\right\} $ were \textit{clinical} psychologists
and $B$ was a \textit{community} psychologist. In this election, voters
ranked the five candidates in order of preference. Among the 15449 votes,
5738 votes ranked all five candidates. We consider the data set which
records the 5738 complete votes; it is available in $\left[ 20,\ p.96\right]
$ and $\left[ 5,\ Table\ 1\right] $. The winner was candidate $C$.

\begin{figure}[h]
\begin{subfigure}{.5\textwidth}
  \centering
  \includegraphics[width=.8\linewidth]{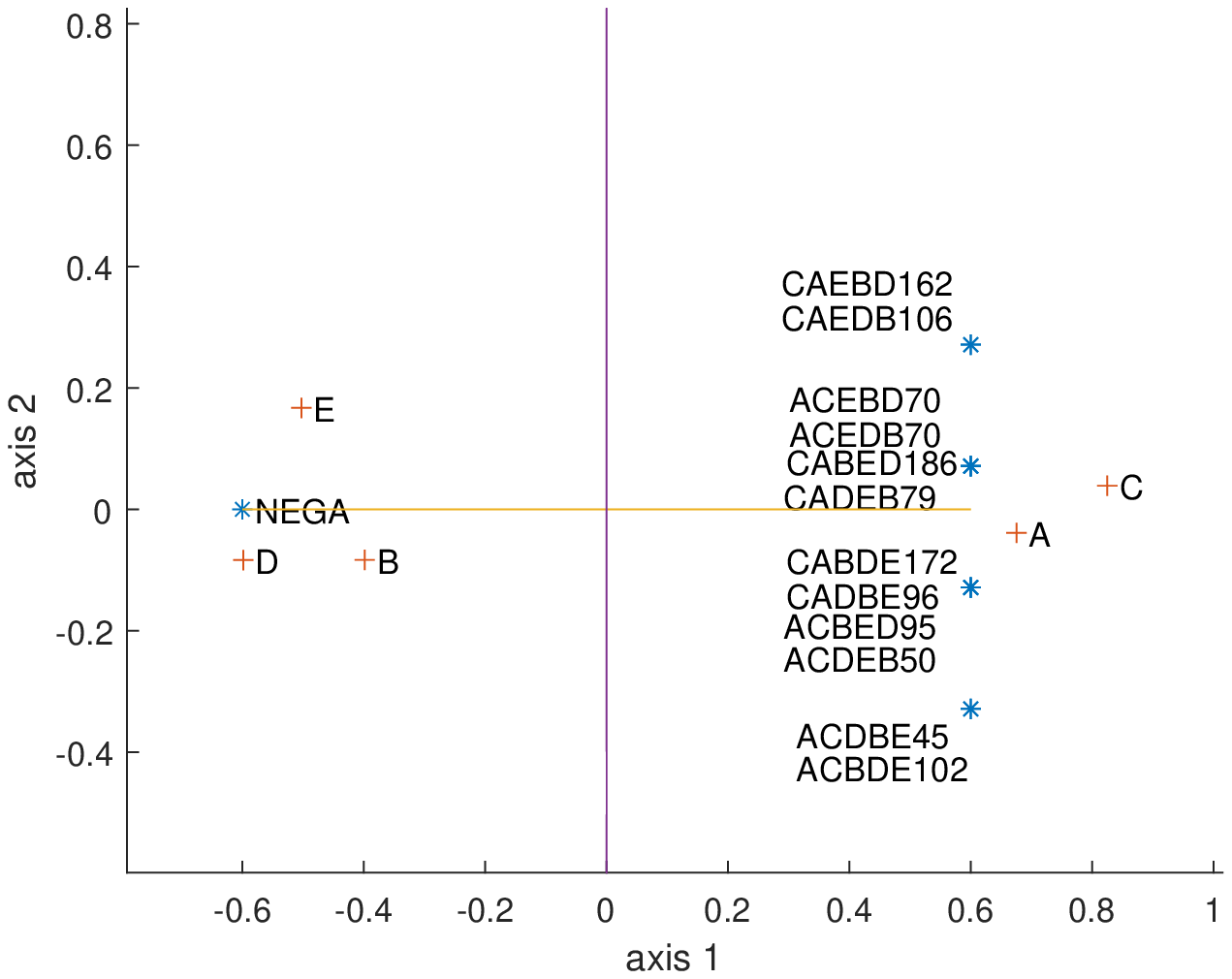}
  \caption{{\tiny Figure 11 :  TCA map of $\mbox{Coh}_{1}C_{1}$ of APA data}}
  \label{fig:sub-third}
\end{subfigure}
\begin{subfigure}{.5\textwidth}
  \centering
  \includegraphics[width=.8\linewidth]{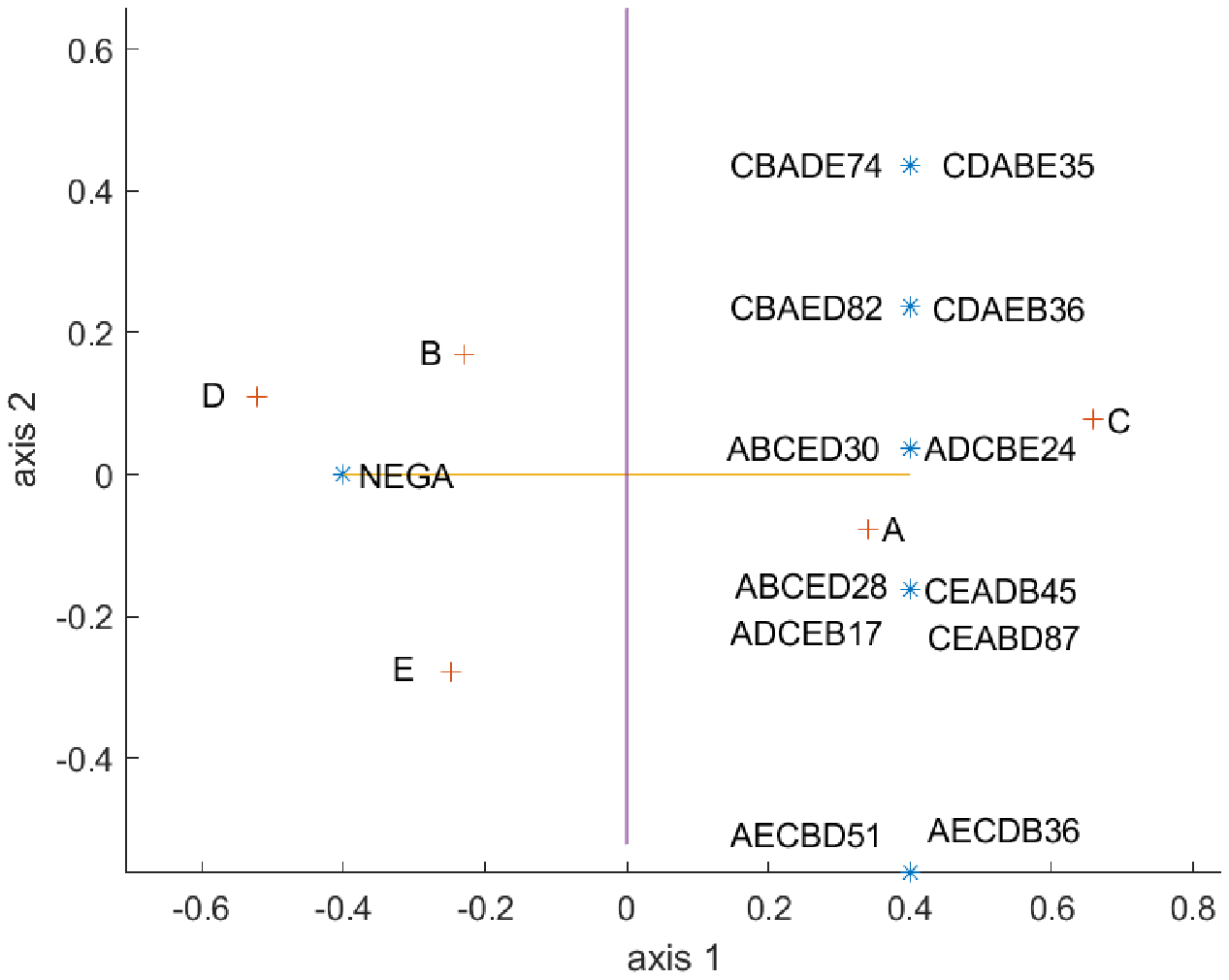}
  \caption{{\tiny Figure 12:  TCA map of $\mbox{Coh}_{1}C_{2}$ of APA data}}
  \label{fig:sub-fourth}
\end{subfigure}
\end{figure}

Table 15 compares the results obtained by our method and the best
distance-based mixture model given in $\left[ 21\right] $. Distance-based
models have two parameters, a central modal ranking and a precision
parameter. The precision parameter measures the peakedness of the
distribution. $\left[ 21\right] $ found that the Cayley distance produced
better results than the Kendall and Spearman distances using BIC (Bayesian
information criterion) and ICL (integrated complete likelihood) criteria.
Parts a and b of Table 15, are reproduced from $\left[ 21,\ \text{Tables 4
and 5}\right] $.

Part c of Table 15 summarizes the results of our approach, where we only
describe the first four coherent groups: We find only the first two coherent
groups as meaningfully interpretable based on the a priori knowledge of the
candidates. Voters in $cohG(1)$, with sample size of 31\%, prefer the
research oriented psychologists $\left\{ A,C\right\} $ over the rest. Voters
in $cohG(2)$, with sample size of 23.7\%, prefer the clinical psychologists $%
\left\{ D,E\right\} $ over the rest. We interpret $cohG(3)$ and $cohG(4)$ as
mixed B with 14.23\% and 12.\% of the voters, respectively. Additionally,
there is a $noisyG$ making up 19.1\% of the sample, which comprises $cohG(5)$
displayed in Table 15.

$\left[ 5\right] $ discussed this data set quite in detail; surprisingly,
our results confirm his observations: a) There are two groups of candidates,
$\left\{ A,C\right\} $ and $\left\{ D,E\right\} .$ The voters line up behind
one group or the other ; b) The APA divides into academicians and clinicians
who are on uneasy terms. Voters seem to choose one type or the other, and
then choose within; but the group effect predominates ; c) Candidate $B$
seems to fall in the middle, perhaps closer to $D$ and $E$.

The following important observation emerges from the comparison of results
in Table 15. We have two distinct concepts of groups for rank data,
categorical and latent variable based. To see this, consider groups 3 and 4
in part a of Table 15: Group 3 is based on the modal category $B\succ C\succ
A\succ D\succ E$ and group 4 is based on the modal category $B\succ C\succ
A\succ E\succ D.$ The only difference between these two modal categories is
the permutation of the least ranked two clinical psychologist candidates $%
\left\{ D,E\right\} ;$ this difference is not important and does not appear
in our approach, which is a latent variable approach.

{\tiny\begin{tabular}{|l|r||r|r|r|r|r||r||}
\multicolumn{8}{l}{\textbf{Table 15: A summary of results derived from three
methods of analysis of}} \\
\multicolumn{8}{l}{\textbf{APA\ election data. Parts a) and b) are from
Murphy and Martin (2003).}} \\ \hline\hline
\multicolumn{8}{|l}{\textit{a) Parameters of the best mixture model
selected, Cayley-based, using BIC}} \\ \hline
Group & sample\% & \multicolumn{5}{||r}{modal orderings} & precision \\
\hline
$1$ & $42$ & \multicolumn{5}{||r}{$D\succ B\succ E\succ C\succ A$} & $0.16$
\\
$2$ & $31$ & \multicolumn{5}{||r}{$C\succ D\succ E\succ A\succ B$} & $0.79$
\\
$3$ & $12$ & \multicolumn{5}{||r}{$B\succ C\succ A\succ D\succ E$} & $1.52$
\\
$4$ & $8$ & \multicolumn{5}{||r}{$B\succ C\succ A\succ E\succ D$} & $1.81$
\\
$5$ & $7$ & \multicolumn{5}{||r}{$B\succ D\succ A\succ E\succ C$} & $1.72$
\\ \hline\hline
\multicolumn{8}{|l}{\textit{b) Parameters of the best mixture model
selected, Cayley-based, using ICL}} \\ \hline
Group & sample\% & \multicolumn{5}{||r}{modal ordering} & precision \\ \hline
$1$ & $100$ & \multicolumn{5}{||r}{$B\succ C\succ A\succ E\succ D$} & $0.25$
\\ \hline\hline
\multicolumn{8}{|l||}{\textit{c) The first five coherent groups, each
composed of two coherent clusters.}} \\ \hline
Group & sample\% & $\beta (C)$ & $\beta (A)$ & $\beta (B)$ & $\beta (E)$ & $%
\beta (D)$ & $Cross$ \\ \hline
cohG(1) Research & $31.0$ & $\mathbf{3.55}$ & $\mathbf{3.15}$ & $1.31$ & $%
1.15$ & $0.85$ & $10.22\%$ \\
cohG(2) Clinical & $23.7$ & $0.83$ & $1.28$ & $1.28$ & $\mathbf{3.31}$ & $%
\mathbf{3.30}$ & $12.90\%$ \\
cohG(3) mixed B & $14.2$ & $0.66$ & $\mathbf{2.70}$ & $\mathbf{2.96}$ & $%
0.71 $ & $\mathbf{2.97}$ & $12.45\%$ \\
cohG(4) mixed B & $12.0$ & $\mathbf{2.85}$ & $0.77$ & $\mathbf{2.86}$ & $%
\mathbf{2.80}$ & $0.72$ & $10.22\%$ \\ \hline
cohG(5) outlier & $8.6$ & $0.96$ & $\mathbf{3.30}$ & $1.31$ & $\mathbf{3.40}$
& $1.00$ & $9.88\%$ \\ \hline
\end{tabular}}

\subsection{Description}

The eight coherent clusters of the first four coherent groups can simply be
described as:

$coh_{1}C(1):T_{v}(\tau _{J_{2}}(S_{2})=\tau _{\left\{ A,C\right\} }\left\{
3,4\right\} =\left\{ 3,4\right\} )=7$ for $v=1,...,1233.$

$coh_{1}C(2):T_{v}(\tau _{J_{2}}(S_{2})=\tau _{\left\{ A,C\right\} }\left\{
3,4\right\} =\left\{ \mathbf{2},4\right\} )=6$ for $v=1,...,545.$

$coh_{2}C(1):T_{v}(\tau _{J_{2}}(S_{2})=\tau _{\left\{ D,E\right\} }\left\{
3,4\right\} =\left\{ 3,4\right\} )=7$ for $v=1,...,834.$

$coh_{2}C(2):T_{v}(\tau _{J_{2}}(S_{2})=\tau _{\left\{ D,E\right\} }\left\{
3,4\right\} =\left\{ \mathbf{2},4\right\} )=6$ for $v=1,...,526.$

$coh_{3}C(1):T_{v}(\tau _{J_{1}}(S_{1})=\tau _{\left\{ C,E\right\} }\left\{
0,1\right\} =\left\{ 0,1\right\} )=1$ for $v=1,...,512.$

$coh_{3}C(2):T_{v}(\tau _{J_{1}}(S_{1})=\tau _{\left\{ C,E\right\} }\left\{
0,1\right\} =\left\{ 0,\mathbf{2}\right\} )=2$ for $v=1,...,305.$

$coh_{4}C(1):T_{v}(\tau _{J_{1}}(S_{1})=\tau _{\left\{ A,D\right\} }\left\{
0,1\right\} =\left\{ 0,1\right\} )=1$ for $v=1,...,350.$

$coh_{4}C(2):T_{v}(\tau _{J_{1}}(S_{1})=\tau _{\left\{ A,D\right\} }\left\{
0,1\right\} =\left\{ 0,\mathbf{2}\right\} )=2$ for $v=1,...,338.$

In this case, we can also visualize all the orderings belonging to a
coherent group: Figures 11 and 12 display all the preferences belonging to
the two coherent clusters of the first coherent group. The label $CAEBD162$
in Figure 11 should be interpreted as the preference $C\succ A\succ E\succ
B\succ D$ repeated 162 times.

\section{Riffle independence model}

Riffle independence is a nonparametric probabilistic modelling method of
preferences developed by $\left[ 2\right] $, which generalizes the
independence model. It can be described in the following way:

(a) Partition the set $J$ of $d$ distinct items into two disjoint subsets $%
J_{1}$ of size $d_{1}$ and $J_{2}$ of size $d_{2}$. Then generate an
ordering of items within each subset according to a certain ranking model.
This implies that any ordering of the $d$ items can be written as a direct
product of two disconnected orderings; which in its turn implies the
independence of the two subsets $J_{1}$ and $J_{2}$. So the model complexity
of this step is of order $d_{1}!+d_{2}!.$

(b) Interleave the two independent orderings for these two subsets using a
riffle shuffle to form a combined ordering. An interleaving is a binary
mapping from the set of orderings to $\left\{ J_{1},J_{2}\right\} $. The
model complexity of this step is of order $d!/(d_{1}!d_{2}!).$ The
interleaving step generates the riffled independence of the two subsets $%
J_{1}$ and $J_{2}$.

So the combined model complexity of both steps is $%
d_{1}!+d_{2}!+d!/(d_{1}!d_{2}!)$ which is much smaller than $%
d!=(d_{1}+d_{2})!$.

For example, consider an ordering of the items in the set $J=\left\{
A,B,C,D,E,F\right\} $ from its two subsets $J_{1}=\left\{ A,C\right\} $ and $%
J_{2}=\left\{ B,D,E,F\right\} .\ $In the first step, relative orderings of
the items in $J_{1}$ and $J_{2}$ are drawn independently. Suppose we obtain
the relative ordering $\varphi (J_{1})=(C\succ A)$ in $J_{1},$ and the
relative ordering $\varphi (J_{2})=(B\succ D\succ F\succ E)$ in $J_{2}.$
Then, in the second step, the two relative orderings are combined by
interleaving the items in the two subsets. For instance, if the interleaving
process is $\omega (J_{1},J_{2})=(J_{1},J_{2},J_{2},J_{1},J_{2},J_{2})$,
where the relative ordering of the items in each subset remains unchanged,
the combined ordering is then determined by the composition
\begin{eqnarray*}
\omega (J_{1},J_{2})\ast (\varphi (J_{1}),\varphi (J_{2})) &=&(C\succ B\succ
D\succ A\succ F\succ E) \\
&=&\varphi (J).
\end{eqnarray*}

Given the two subsets $J_{1}$ and $J_{2}$ with their orderings $\varphi
(J_{1})$ and $\varphi (J_{2})$ and interleaving $\omega (J_{1},J_{2})$
generated from models with probability distributions $f_{J_{1}},$ $g_{J_{2}}$
and $m_{\omega }$, respectively, the probability of observed ordering under
the riffle independence model is
\begin{equation*}
P(\varphi (J))=m_{\omega }(\omega (J_{1},J_{2}))f_{J_{1}}(\varphi
(J_{1}))g_{J_{2}}(\varphi (J_{2}).
\end{equation*}%
There are two formulations of riffle shuffle for rank data in statistics:
probabilistic and exploratory. In the riffled independence model, the set of
items is partitioned recursively, while in the exploratory approach the set
of voters is partitioned recursively.

\section{\textbf{Conclusion}}

The main contribution of this paper is the introduction of an exploratory
riffle shuffling procedure to reveal and display the structure of diffuse
rank data for large sample sizes. The new notion of a coherent cluster, that
we developed, is simply based on the geometric notion of taxicab projection
of points on the first TCA axis globally and locally; furthermore, it has
nice mathematical properties. Coherent clusters of a coherent group
represent the same latent variable opposing preferred items to disliked
items, and can easily be interpreted and displayed.

Like Occam's razor, step by step, our procedure peels the essential
structural layers (coherent groups) of rank data.

Our method was able to discover some other aspects of the rank data, such as
outliers or small groups, which are eclipsed or masked by well established
methods, such as distance or random utility based methods. The major reasons
for this is that in random utility based methods the multivariate nature of
a preference is reduced to binary preferences (paired comparisons), and in
Mallows distance related methods distances between any two preferences are
bounded.

We presented a new index, $Cross$, that quantifies the extent of crossing of
scores between the optimal binary partition of the items that resulted from
TCA. The crossing index of a group is based on the first taxicab dispersion
measure: it takes values between 0 and 100\%, so it is easily interpretable.

The proposed approach can easily be generalized to the analysis of rankings
with ties and partial rankings.

The package TaxicabCA written in R available on CRAN can be used to do the
calculations.\bigskip

\textit{Acknowledgement: }Choulakian's research has been supported by NSERC
grant (RGPIN-2017-05092) of Canada.\bigskip

\textbf{References}\medskip

$\left[ 1\right] \ \ $Kamishima, T. (2003). Nantonac collaborative
filtering: recommendation based on order responses. In: \textit{Proceedings
of the ninth ACM SIGKDD international conference on Knowledge discovery and
data mining. KDD '03}, 583--588. ACM, New York.

$\left[ 2\right] \ \ $Huang, J., Guestrin, C. (2012). Uncovering the riffled
independence structure of ranked data. \textit{Electronic Journal of
Statistics, 6,} 199-230.

$\left[ 3\right] \ \ $Lu, T., Boutilier, C. (2014). Effective sampling and
learning for Mallows models with pairwise preference data. \textit{Journal
of Machine Learning Research, 15,} 3783-3829.

$\left[ 4\right] \ \ $Vitelli, V., S\o renson, \O ., Crispino, M., Frigessi,
A., Arjas, E. (2018). Probabilistic preference learning with the Mallows
rank model. \textit{Journal of Machine Learning Research, 18,} 1-49.

$\left[ 5\right] \ \ $Diaconis, P. (1989). A generalization of spectral
analysis with application to ranked data. \textit{The Annals of Statistics,
17}(3), 949-979.

$\left[ 6\right] \ \ $Marden, J.I. (1995). \textit{Analyzing and Modeling of
Rank Data}. Chapman \& Hall, London

$\left[ 7\right] \ \ $Alvo, M., Yu, P. (2014). \textit{Statistical Methods
for Ranking Data}. Springer, New York

$\left[ 8\right] \ \ $Bayer, D., Diaconis, P. (1992). Trailing the dovetail
shuffle to its lair. \textit{The Annals of Probability}, \textit{2(2)},
294-313.

$\left[ 9\right] \ \ $Choulakian, V. (2016). Globally homogenous mixture
components and local heterogeneity of rank data. \textit{arXiv:1608.05058}.

$\left[ 10\right] \ \ $Choulakian, V. (2006). Taxicab correspondence
analysis. \textit{Psychometrika, 71}, 333-345.

$\left[ 11\right] \ \ $Choulakian, V. (2016). Matrix factorizations based on
induced norms. \textit{Statistics, Optimization and Information Computing, 4,%
} 1-14.

$\left[ 12\right] \ \ $Borda, J. de, (1781). M\'{e}moire sur les \'{e}%
lections au scrutin. \textit{Histoire de L'Acad\'{e}mie Royale des Sciences}%
, \textit{102}, 657-665.

$\left[ 13\right] \ \ $Benz\'{e}cri, J.P. (1991). Comment on Leo A.
Goodman's invited paper. \textit{Journal of the American Statistical
Association,} \textit{86}, 1112-1115.

$\left[ 14\right] \ \ $Van de Velden, M. (2000). Dual scaling and
correspondence analysis of of rank order data. In: Heijmans, Pollock,
Satorra (eds), \textit{Innovations in multivariate statistical analysis, 36}%
: 87-99. Kluwer Academic Publishers, Dordrecht.

$\left[ 15\right] \ \ $Torres, A., Greenacre, M. (2002). Dual scaling and
correspondence analysis of preferences, paired comparisons and ratings.
\textit{International Journal of Research in Marketing, 19}(4), 401-405.

$\left[ 16\right] \ \ $Nishisato, S. (1980). Analysis of Categorical Data:
Dual Scaling and Its Applications. Toronto:University of Toronto

Press.

$\left[ 17\right] \ \ $Choulakian, V. (2014). Taxicab correspondence
analysis of ratings and rankings. \textit{Journal de la Soci\'{e}t\'{e} Fran%
\c{c}aise de Statistique, 155}(4), 1-23.

$\left[ 18\right] \ \ $Khot, S. and Naor, A. (2012). Grothendieck-type
inequalities in combinatorial optimization. \textit{Communications on Pure
and Applied Mathematics, Vol. LXV}, 992-1035.

$\left[ 19\right] \ \ $Choulakian, V., Abou-Samra, G. (2020). Mean absolute
deviations about the mean, cut norm and taxicab correspondence analysis.
\textit{Open Journal of Statistics, 10}(1), 97-112.

$\left[ 20\right] \ \ $Diaconis, P. (1988). \textit{Group Representations in
Probability and Statistics}. Institute of Mathematical Statistics, Hayward,
CA.

$\left[ 21\right] \ \ $Murphy, T.B., Martin, D. (2003). Mixtures of
distance-based models for ranking data. \textit{Computational Statistics and
Data Analysis, 41}, 645-655.

\bigskip

\textbf{Appendix\bigskip }

Let $\mathbf{R}=(r_{ij})$ for $i=1,...,n$ and $j=1,...,d$ represent the
Borda scorings for preferences, where $r_{ij}$ takes values $0,...,d-1.$
Similarly, let $\overline{\mathbf{R}}$ represent the reverse Borda scorings,
whose column sums are the cordinates of the row named $\mathbf{nega=}$ $n%
\overline{\mathbf{\beta }}=\mathbf{1}_{n}^{\prime }\overline{\mathbf{R}}.$
We consider the application of TCA to the data set
\begin{equation*}
\mathbf{R}_{nega}=(_{\mathbf{nega}}^{\mathbf{R}})
\end{equation*}%
of size $(n+1)\times d.$ So let
\begin{equation*}
\mathbf{P}=\mathbf{R}_{nega}/t
\end{equation*}%
be the correspondence table associated with $\mathbf{R}_{nega},$ where $%
t=2n\sum_{j=0}^{d-1}=nd(d-1).$ We have

\begin{equation}
\begin{array}{cccc}
p_{i\ast } &=&\frac{1}{2n}\;\;\;\text{for}\quad i=1,...,n&  \tag{14}
\end{array}
\end{equation}
\begin{equation}
\begin{array}{cccc}
 &=&\frac{1}{2}\;\;\;\text{for}\quad i=n+1,& \tag{15}
\end{array}
\end{equation}
and

\begin{equation}
p_{\ast j}=\frac{1}{d}\text{\ \ \ \ \ for\ \ \ }j=1,...,d. \tag{16}
\end{equation}%
The first residuel correspondence matrix will be%

\begin{equation*}
\begin{array}{ccccc}
p_{ij}^{(1)} &=&p_{ij}-p_{i\ast }p_{\ast j} & \qquad \qquad \qquad (17)\\
&=&\frac{r_{ij}}{t}-\frac{1}{2n}.\frac{1}{d}\text{\ \ \ \ for\ \ \ \ }%
i=1,...,n & \qquad \qquad \qquad (18)\\
&=&\frac{\mathbf{nega}_{j}}{t}-\frac{1}{2}.\frac{1}{d}\ \ \ \ \text{for\ \ \
\ }i=n+1.  & \qquad \qquad \qquad (19)
\end{array}
\end{equation*}

Consider the nontrivial binary partition of the set $S=\left\{
0,1,...,d-1\right\} $ into $S=S_{1}\cup S_{2},$ where $|S_{1}|=d_{1},$ $%
|S_{2}|=d_{2}$ and $d=d_{1}+d_{2}.$ To eliminate the sign indeterminacy in
the first TCA principal axis, we fix $\mathbf{v}_{1}(nega)=\mathbf{v}%
_{1}(n+1)=-1;$ and we designate by $S_{1}$ the set of item indices such that
the first TCA principal axis coordinates are negative, that is, $\mathbf{u}%
_{1}(j)=-1$\ \ for $j\in S_{1}.$ It follows that $\mathbf{u}_{1}(j)=1$\ \
for $j\in S_{2}$.

Now we have by (4) for $i=1,...,n$

\begin{equation}
\begin{array}{llll}
a_{i1} &=&\sum_{j=1}^{d}\mathbf{u}_{1}(j)p_{ij}^{(1)}  \notag \\
&=&\sum_{j\in S_{1}}\mathbf{u}_{1}(j)p_{ij}^{(1)}+\sum_{j\in S_{2}}\mathbf{u}%
_{1}(j)p_{ij}^{(1)} & \notag \\
&=&-\sum_{j\in S_{1}}p_{ij}^{(1)}+\sum_{j\in S_{2}}p_{ij}^{(1)} \\
&=&-2\sum_{j\in S_{1}}p_{ij}^{(1)}\ \ \ \ \text{by\ \ }(17)& \notag \\
&=&-2\sum_{j\in S_{1}}(\frac{r_{ij}}{t}-\frac{1}{2n}.\frac{1}{d})\text{\ \ \
by\ \ (18)} & \notag \\
&=&\frac{d_{1}}{nd}-\frac{2}{t}\sum_{j\in S_{1}}r_{ij}; & \tag{20}
\end{array}
\end{equation}

and from which we deduce by (5) for $i=1,...,n$
\begin{equation}
\begin{array}{lll}
f_{i1} &=&\frac{a_{i1}}{p_{i\ast }}  \notag \\
&=&\frac{2d_{1}}{d}-\frac{4}{d(d-1)}\sum_{j\in S_{1}}r_{ij}.  \tag{21}
\end{array}
\end{equation}%
\bigskip We have the following Theorem concerning the first TCA principal
factor scores of respondents $f_{i1}$ for $i=1,...,n.$

\textbf{Theorem 1}:

a) The maximum number of distinct clusters of $n$ respondents on the first
TCA principal axis (distinct $f_{i1}$ values$)$ is $d_{1}d_{2}+1.$

\textit{Proof}: We consider the two extreme cases of $S_{1}$ and calculate
the summation term in (21):

For $S_{1}=\left\{ 0,1,...,d_{1}-1\right\} $,$\ \sum_{j\in
S_{1}}r_{ij}=\sum_{j=0}^{d_{1}-1}j=\frac{d_{1}(d_{1}-1)}{2}.$

For $S_{1}=\left\{ d-d_{1},1,...,d-1\right\} ,$ $\sum_{j\in
S_{1}}r_{ij}=\sum_{j=d-d_{1}}^{d-1}j=\sum_{j=d_{2}}^{d-1}j=\frac{%
d_{1}(d_{2}+d-1)}{2}$.

It follows that
\begin{equation*}
\frac{d_{1}(d_{1}-1)}{2}\leq \sum_{j\in S_{1}}r_{ij}\leq \frac{%
d_{1}(d_{2}+d-1)}{2};
\end{equation*}
so $\sum_{j\in S_{1}}r_{ij}$ can take at most $\frac{d_{1}(d_{2}+d-1)}{2}-%
\frac{d_{1}(d_{1}-1)}{2}+1=d_{1}d_{2}+1$ values.

b) The maximum value that $f_{i1}$ can attain is $2\frac{d_{1}d_{2}}{d(d-1)}%
. $

\textit{Proof}: From (21) and Part a, it follows that the maximum value that
$f_{i1}$ can attain is $(\frac{2d_{1}}{d}-\frac{4}{d(d-1)}\frac{%
d_{1}(d_{1}-1)}{2})=2\frac{d_{1}d_{2}}{d(d-1)}.$

c) The minimum value that $f_{i1}$ can attain is $-2\frac{d_{1}d_{2}}{d(d-1)}%
.$

\textit{Proof}: From (21) and Part a, it follows that the minimum value that
$f_{i1}$ can attain is $(\frac{2d_{1}}{d}-\frac{4}{d(d-1)}\frac{%
d_{1}(d_{2}+d-1)}{2})=-2\frac{d_{1}d_{2}}{d(d-1)}.$

d) If the number of distinct clusters is maximum, $d_{1}d_{2}+1$, then the
gap between two contiguous $f_{i1}$ values is $\frac{4}{d(d-1)}.$

\textit{Proof}: Suppose that the number of distinct clusters is maximum, $%
d_{1}d_{2}+1$. We consider the first TCA factor score $f_{i1}=\frac{2d_{1}}{d%
}-\frac{4}{d(d-1)}\sum_{j\in S_{1}}r_{ij}$ which is different in value from
the two extreme values $\pm 2\frac{d_{1}d_{2}}{d(d-1)}.$ Then $f_{i_{1}1}=%
\frac{2d_{1}}{d}-\frac{4}{d(d-1)}(-1+\sum_{j\in S_{1}}r_{ij})$ will be the
contiguous higher value to $f_{i1};$ and similarly $f_{i_{2}1}=\frac{2d_{1}}{%
d}-\frac{4}{d(d-1)}(1+\sum_{j\in S_{1}}r_{ij})$ will be the contiguous lower
value to $f_{i1};$ and the required result follows.\bigskip

\textbf{Proposition 1:} For a voting profile $V$, $\delta _{1}\geq |f_{1}(%
\mathbf{nega})|$.

\textit{Proof}: Let $\mathbf{a}_{1}=(_{a_{1}(nega)}^{\mathbf{a}_{11}}).$We
need the following three observations.

First, it is well known that $\mathbf{a}_{1}$ is centered by (5) and (9),
\begin{eqnarray*}
\mathbf{1}_{n+1}^{\prime }\mathbf{a}_{1} &=&0, \\
&=&\mathbf{1}_{n}^{\prime }\mathbf{a}_{11}+a_{1}(nega);
\end{eqnarray*}%
from which we get,%
\begin{equation}
|\mathbf{1}_{n}^{\prime }\ \mathbf{a}_{11}|=|a_{1}(nega)|.  \tag{22}
\end{equation}%
Second, by triangle inequality of the L$_{1}$ norm we have%
\begin{equation}
||\mathbf{a}_{11}||_{1}\geq |\mathbf{1}_{n}^{\prime }\mathbf{a}_{11}|.
\tag{23}
\end{equation}

Third, the marginal relative frequency of the nega row is $p_{nega\ast }=1/2$
by (15) , and $f_{i1}=a_{i1}/p_{i\ast }$ for $i=1,...,n+1$ by (5); so we
have
\begin{equation}
f_{1}(nega)=2a_{1}(nega).  \tag{24}
\end{equation}

Now we have by (7)
\begin{equation}
\begin{array}{llll}
\delta _{1} &=&||\mathbf{a}_{1}\mathbf{||}_{1}  \notag \\
&=&||\mathbf{a}_{11}||_{1}+|a_{1}(nega)|  \notag \\
&\geq &|\mathbf{1}_{n}^{\prime }\mathbf{a}_{11}|+|a_{1}(nega)|\text{\ \ \ by
\ \ (23)} \tag{25} \\
&=&2|a_{1}(nega)|\ \ \ \ \text{by\ \ \ (22)}  \notag \\
&=&|f_{1}(nega)|\text{\ \ \ \ by\ \ (24)}   \notag
\end{array}
\end{equation}%
\bigskip

\textbf{Propostion 2}: Let $cohC_{m}(\alpha )=V_{m,\alpha }$ be the $\alpha $%
th coherent cluster of the $m$th coherent group characterized by $%
f_{1}^{V_{m,\alpha }}(\mathbf{\sigma )=}f_{\alpha }^{V_{m}}$ for all $%
\mathbf{\sigma \in }cohC_{m}(\alpha )$. Then $\delta _{1}=f_{\alpha
}^{V_{m}}=-f_{1}(\mathbf{nega}).$

\textit{Proof}: By Definition 1 of the coherency of the cluster $V_{m,\alpha
},$ we have $0<f_{1}^{V_{m,\alpha }}(i)=f_{\alpha }^{V_{m}}$ for $%
i=1,...,|cohC_{m}(\alpha )|$; by (5) it follows that $0<a_{i1}=$ $f_{\alpha
}^{V_{m}}/n$ for $i=1,...,|cohC_{m}(\alpha )|$ ; so (25) becomes equality, $%
||\mathbf{a}_{11}||_{1}=\sum_{i=1}^{n}a_{i1}=|\mathbf{1}_{n}^{\prime }%
\mathbf{a}_{11}|$, and the required result follows.\bigskip

\textbf{Proposition 3} is a corollary to the following general result\bigskip

\textbf{Theorem 3}: If the first TCA principal axis of the columns of $%
\mathbf{R}_{nega}$ is\ $\mathbf{v}_{1}=(_{-1}^{\mathbf{1}_{n}})$, then

the first principal column factor score $\mathbf{g}_{1}$ of the $d$ items is
an affine function of the Borda scale $\mathbf{\beta };$ that is, $g_{1}(j)=%
\frac{2}{d-1}\beta (j)-1$\ \ or\ \ $corr(\mathbf{g}_{1},\mathbf{\beta })=1.$

\textit{Proof}: Suppose that $\mathbf{v}_{1}=(_{-1}^{\mathbf{1}_{n}});$ then
by (4) for $j=1,...,d$
\begin{eqnarray*}
b_{1}(j) &=&\sum_{i=1}^{n+1}v_{1}(i)p_{ij}^{(1)} \\
&=&\sum_{i=1}^{n}p_{ij}^{(1)}-p_{(n+1)j}^{(1)} \\
&=&2\sum_{i=1}^{n}p_{ij}^{(1)}\text{ \ \ by\ \ \ (17)} \\
&=&2\sum_{i=1}^{n}(p_{ij}-p_{i\ast }p_{\ast j}) \\
&=&2\sum_{i=1}^{n}r_{ij}/t-p_{\ast j}\ \ \ \text{by\ \ (14)} \\
&=&2n\beta (j)/t-p_{\ast j}
\end{eqnarray*}%
Thus by (5) for $j=1,...,d$
\begin{eqnarray*}
g_{1}(j) &=&b_{1}(j)/p_{\ast j} \\
&=&\frac{2n\beta (j)/t-p_{\ast j}}{p_{\ast j}} \\
&=&\frac{2\beta (j)}{d-1}-1.
\end{eqnarray*}

\textbf{Proposition 4}: The crossing index of a coherent cluster is%
\begin{equation*}
Cross(cohC(\alpha ))=\frac{2(\alpha -1)}{d_{1}d_{2}}.
\end{equation*}

\textit{Proof: }Easily shown by using Definition 3 and Proposition 2.\bigskip

The proof of Theorem 2a easily follows from Theorem 3. The proof of Theorem
2b is similar to the proof of Propostion 1. The proof of Theorem 2c is
similar to the proof of Propostion 4.

\end{document}